\documentclass[prb,aps,amsmath,amssymb,twocolumn,groupedaddress,floats,showpacs,final,reprint]{revtex4-1}
\usepackage{graphicx}
\usepackage{amssymb,amsmath}
\usepackage{color}
\usepackage{bm}

\newcommand{\melement}[3]{\left<#1\left|\vphantom{#1}#2\vphantom{#3}\right|#3\right>}		
\newcommand{\braket}[2]{\left<\left.#1\vphantom{#2}\right|#2\right>}				
\newcommand{\bra}[1]{\left<#1\right|}								
\newcommand{\ket}[1]{\left|#1\right>}								
\newcommand{\abs}[1]{\left|#1\right|}								

\begin{document}

\author{Peter Kroiss}
\author{Lode Pollet}
\affiliation{Department of Physics, Arnold Sommerfeld Center for Theoretical Physics and Center for NanoScience, University of Munich, Theresienstrasse 37, 80333 Munich, Germany}

\title{Diagrammatic Monte Carlo study of mass-imbalanced Fermi-polaron system}


\date{\today}

\begin{abstract}
We apply the diagrammatic Monte Carlo approach to three-dimensional Fermi-polaron systems with mass-imbalance, where an impurity interacts resonantly with a noninteracting Fermi sea whose atoms have a different mass.
This method allows to go beyond frequently used variational techniques by stochastically summing all relevant impurity Feynman diagrams up to a maximum expansion order limited by the sign problem. 
Polaron energy and quasiparticle residue can be accurately determined over a broad range of impurity masses. Furthermore, the spectral function of an imbalanced polaron demonstrates the stability
of the quasiparticle and allows to locate in addition also the repulsive polaron as an excited state. 
The quantitative exactness of two-particle-hole wave-functions is investigated, resulting in a relative lowering of polaronic energies
in the mass-imbalance phase diagram.
Tan's contact coefficient for the mass-balanced polaron system is found in good agreement with variational methods. 
Mass-imbalanced systems can be studied experimentally by ultracold atom mixtures like $^6$Li--$^{40}$K. 
\end{abstract}


\maketitle

\section{Introduction}

One of the most general and successful concepts in physics is the separation of a physical system into a simpler, controlled subsystem that is interacting with a perturbing subsystem.
A specific example of this method is given by a basic impurity problem, consisting of a noninteracting homogeneous medium and one particle disturbing it.
In the case of a noninteracting Fermi gas, this is called Fermi-polaron problem\cite{massignan2014}.
This theoretical model can help to map out the phase diagram of a strongly population-imbalanced Fermi gas\cite{chevy2006}, where the 
quasiparticle energy and effective mass serve as input parameters for Landau-Pomeranchuk Hamiltonians\cite{pilati2008,parish2011} helping to quantify zero temperature phase separation
and the ground state energy of different phases.
Moreover, the $N+1$ Fermi-polaron system was shown to undergo a transition of its own, featuring as possible ground states\cite{prokofev2008A,prokofev2008B}
a polaronic spin-1/2 quasiparticle and the composite spin-0 molecule, consisting of the impurity and a single bath
atom. However, this transition does not generalize easily to population-imbalanced Fermi gases as it might be preempted by phase separation.

Two of the most common approaches to the Fermi-polaron problem are variational ans\"atze
and the use of an approximate diagrammatic technique\cite{chevy2006,combescot2007,combescot2008,punk2009,chevy2009,massignan2011,trefzger2013,combescot2009,baarsma2012}.
These methods are able to calculate key quasiparticle properties, e.g., effective mass or polaron residue and helped to map the transition between polaronic and molecular states.
Mathy {\it et al.}~extended\cite{parish2011} these ideas to the case of a mass-imbalanced Fermi-polaron problem and depicted the ground state phase diagram 
with respect to polaronic, molecular, trimer\cite{kartavtsev2007} and tetramer\cite{blume2012,levinsen2013}
states, where a trimer (tetramer) is the bound state of two (three) bath particles with the impurity.  
Concerning other techniques, Functional Renormalization Group\cite{schmidt2011} and fixed-node diffusion Monte Carlo\cite{lobo2006}
have been successfully applied.
Experimental works include Refs.~\onlinecite{schirotzek2009, nascimbene2009, kohstall2012}.

Another approach was presented by Prokof'ev and Svistunov in 2008, diagrammatic Monte Carlo\cite{prokofev2008B} (diagMC).
Its key ingredient is the sampling of Feynman diagram integrals by a set of ergodic updates linking all topologies and internal variables.
By reducing the diagrammatic space, they managed to reach sufficiently high expansion orders allowing them to extract energies and effective masses in 
very good agreement with other techniques despite the fermionic sign problem.
Recently, the method was used\cite{vlietinck2013,vlietinck2014,kroiss2014} for the extraction of polaron quasiparticle residues and two-dimensional geometries.

Up to now, these diagMC implementations have only been applied to the special case of equal masses of impurity and bath atoms. 
This is important as such a system can be created by different spin states of a homogeneous atomic gas.
However, also mixtures like $^6$Li--$^{40}$K are experimentally realizable in ultracold atom systems. 
In our work, we extend diagMC to the case of arbitrary mass-imbalance and present the dependence
of polaron energy and residue on the imbalance ratio.
Determining the polaronic spectral function for mass-imbalance helps understanding the stability of quasiparticles close to the limit of a heavy impurity. 
It features the repulsive polaron\cite{cui2010}, an excited state with finite lifetime.
We show that two-particle-hole wave-functions remain essentially exact in three dimensions and demonstrate the implications for the mass-imbalanced phase diagram.
We also present results for Tan's contact parameter for a mass-balanced polaron system. 

This paper is structured as follows:
Section \ref{sec:sec2} presents the basic Fermi-polaron model summarizing the diagrammatic ingredients for imbalanced masses.
In Sec.~\ref{sec:sec3}, some changes of the diagrammatic routine are proposed in order to increase performance, while
Sec.~\ref{sec:sec4} introduces the enhancements of bold diagMC we use in our code.
Section \ref{sec:sec5} will explain a newly developed regrouping technique helping to increase extrapolation speed of the series.
Finally, Sec.~\ref{sec:sec6} exhibits our results.
A brief conclusion is given in Sec.~\ref{sec:sec7}, while the appendices give a detailed derivation of the mass-imbalanced $T$ matrix 
and explain our extrapolation procedure.

\section{Model}
\label{sec:sec2}
The system referred to as a Fermi-polaron problem consists of two main ingredients: a noninteracting Fermi bath and an impurity resonantly interacting with it.
This can be realized at ultracold temperatures because s-wave scattering between any bath particles is forbidden
due to Pauli's principle and p-wave scattering is energetically suppressed.
The simplest Hamiltonian compatible with these requirements is~\cite{chevy2006}
\begin{equation}
\hat{H} = \sum_{{\bf k},\sigma} \epsilon_{k,\sigma} \hat{c}_{{\bf k},\sigma}^{\dagger}
\hat{c}_{{\bf k},\sigma} + g \sum_{{\bf k},{\bf k'},{\bf q}}
\hat{c}_{{\bf k}+{\bf q},\uparrow}^{\dagger}\hat{c}_{{\bf k'}-{\bf q},\downarrow}^{\dagger}
\hat{c}_{{\bf k'},\downarrow} \hat{c}_{{\bf k},\uparrow}.
\end{equation}
\(\hat{c}_{{\bf k},\sigma}\) labels a field operator for a particle in state $\sigma$ and with momentum
\({\bf k}\), \(g\) is the bare coupling constant and
\(\epsilon_{k,\sigma} = \frac{k^2}{2m_\sigma}\) incorporates the energy dispersion.
For convenience, we label the impurity by $\downarrow$ and bath particles by $\uparrow$, although this does not necessarily refer to pure spin states.
$k_F$ is the bath particle Fermi momentum, $E_F$ its Fermi energy.
We are working in units assuring $\hbar = 1$. For the diagrammatic series, we employ the imaginary time representation and
establish the following particle-hole convention: Bath lines with momentum $|{\bf k}| > k_F$ are denoted particles possessing positive time
while lines with $|{\bf k}| < k_F$ are called holes and propagate with negative time.
The positive time direction is defined to be from left to right.
The inter-particle interaction can be modeled by a pseudo-potential in case of a zero-range interaction.
It is important, though, that this potential assures the correct two-particle scattering length\cite{castin2007,zwerger2008} $a$.
We introduce the $T$ matrix $\Gamma(\tau,{\bf p})$ in conventional form\cite{prokofev2008B} and tabulate it prior to the Monte Carlo run.
This is tantamount to replacing the bare coupling $g$ by the experimentally accessible scattering length $a$.
We refer to appendix \ref{sec:appA} for details.
The Green's functions are given by 
\begin{equation}
\begin{aligned}
 G^0_{\uparrow}(\tau,{\bf k}) =  
 &- \theta(\tau) \theta(k-k_F) e^{- (\epsilon_{k,\uparrow}-E_F) \tau} \\
 &+ \theta(-\tau) \theta(k_F-k) e^{- (\epsilon_{k,\uparrow}-E_F) \tau} \\
G^0_{\downarrow}(\tau,{\bf k}) =  
 &- \theta(\tau) e^{- (\epsilon_{k,\downarrow}-\mu_{\downarrow}^{0}) \tau},
\end{aligned}
\end{equation}
where $\mu_{\downarrow}^0$ is a tuning parameter used for convergence reasons.
Reduced mass $m_r$ and total mass $M$ are introduced canonically
\begin{equation}
\begin{aligned}
M &= m_\downarrow + m_\uparrow \\
m_r &= \frac{m_\downarrow m_\uparrow}{M}.
\end{aligned}
\end{equation}

\section{Diagrammatic framework}
\label{sec:sec3}
%
\begin{figure}[ptb]
\includegraphics[width=0.3\linewidth]{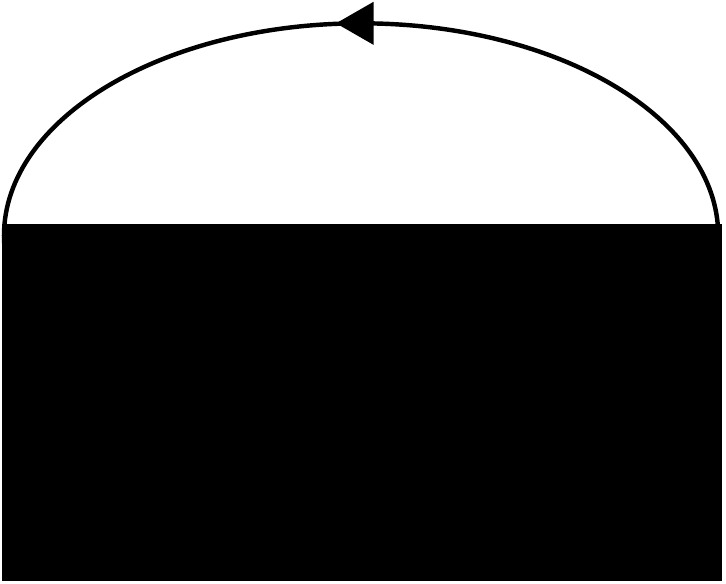}
\caption{\label{fig:fake}  
The first-order diagram is used for normalization purposes.
The (local) appearance of this diagram topology as part of the whole diagram
will be used to identify reducible diagrams in partially bold diagMC in Sec.~\ref{sec:sec4}.}
\end{figure}
%
\begin{figure}[ptb]
\includegraphics[width=0.8\linewidth]{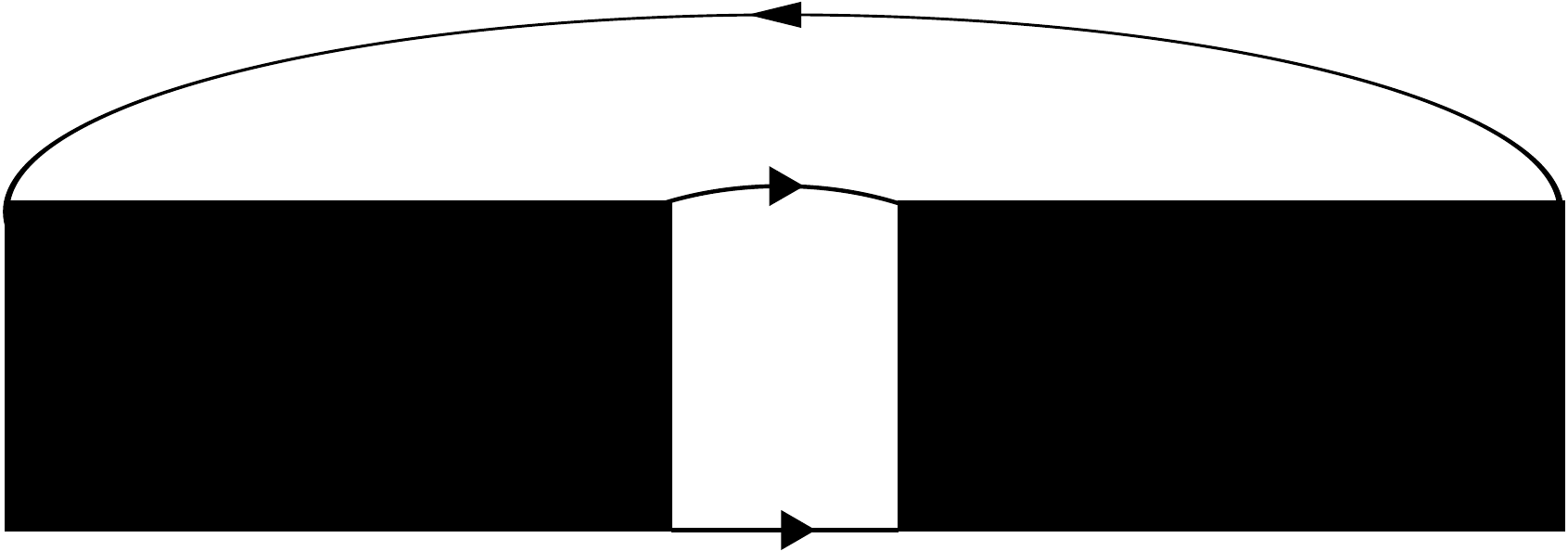}
\caption{\label{fig:order2}  
This (unphysical) second-order diagram connects the first-order fake diagram with higher-order diagrams.}
\end{figure}

The set of updates we use is different from the approach of Prokof'ev-Svistunov\cite{prokofev2008B}. 
Rather than linking different orders by worm diagrams, we prefer to implement these transitions by direct updates.
We propose the following update pairs:
\begin{itemize}
  \item First-to-fake and Fake-to-first,
  \item Change-fake (self-inverse),
  \item Insert-mushroom and Remove-mushroom,
  \item Insert-$T$ matrix and Remove-$T$ matrix,
  \item Reconnect (self-inverse).
\end{itemize}

A fake diagram is used for normalization purposes and is graphically identical to the first-order\cite{order} diagram, but with analytically easy weights, cf.~Fig.~\ref{fig:fake}. Its internal variables are updated by
the update Change-Fake. The updates First-to-fake and Fake-to-first connect this fake diagram with the lowest-order diagram we sample, presented in Fig.~\ref{fig:order2}. 
Note that this diagram is unphysical if no self-consistent bold scheme\cite{prokofev2008B} is used. The first-order diagram is not included in our simulation because its contribution experiences a 
$\frac{1} {\sqrt{\tau}}$-behavior for $\tau \rightarrow 0$, thus forcing the program to spend a lot of time on small times. 
It is straightforward to include the first-order self-energy by a numerical tabulation in $\omega$ space\cite{combescot2007}.

There are four updates linking different orders: Insert-mushroom, Insert-$T$ matrix and their inverse updates Remove-mushroom and Remove-$T$ matrix.
Insert-$T$ matrix chooses any $T$ matrix of the current diagram and splits it into two linked $T$ matrices. The resulting (unphysical) diagram has good overlap with the previous configuration if
$G^1_\downarrow - G^0_\downarrow$ is artificially attributed as weight of the underlying impurity propagator. 
Here, $G^1_\downarrow$ denotes the impurity Green's function evaluated by plugging the first-order self-energy contribution into Dyson's equation.

Last, an update called Reconnect ensures that all different topologies of a certain order are sampled.

This set of updates is ergodic and avoids sampling of first-order contributions.
The last remaining unphysical diagrams connect two adjacent $T$ matrices -- however, this is important for partially bold sampling (cf.~section \ref{sec:sec3}).
If no self-consistent bold scheme is used, sampling of relevant diagrams can be enforced by assigning an additional penalty weight to those diagrams.
We will present the updates Insert-mushroom, Remove-mushroom and Reconnect in the next subsections. All other updates were designed in the same spirit.

\subsection{Insert-mushroom}
\begin{figure}[b]
\includegraphics[width=0.99\linewidth]{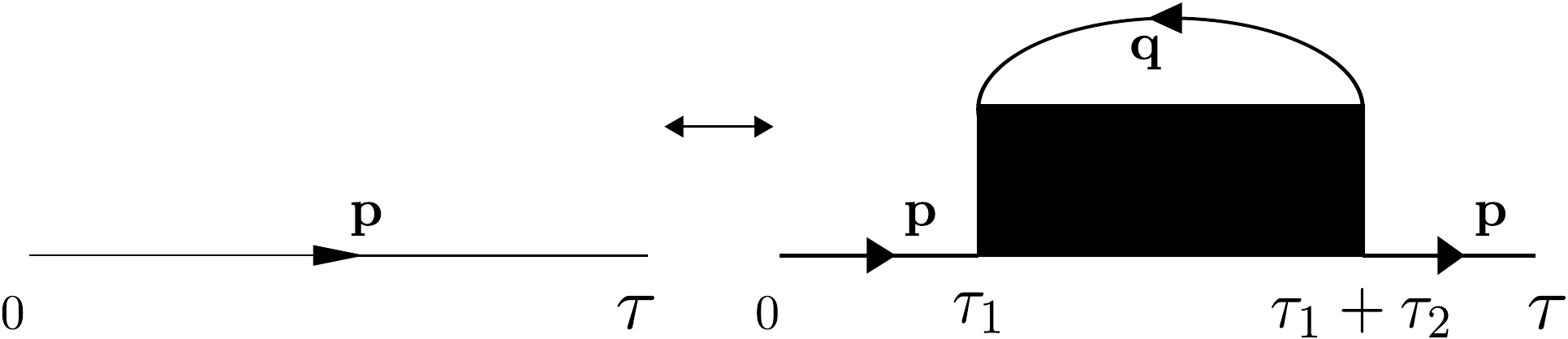}
\caption{\label{fig:insert_mushroom}  
Illustration of inverse updates Insert-mushroom and Remove-mushroom.}
\end{figure}

This update is available for impurity propagator lines.
It attempts to insert the diagrammatic structure of Fig.~\ref{fig:fake} (called mushroom) on one of these lines.
If the current diagram order is denoted by $N$, there are $N-1$ propagators available for insertion.
Having selected one of those with imaginary time $\tau$ and momentum ${\bf p}$, internal time slices $\tau_1$ and $\tau_2$ are uniformly seeded (probabilities: 
$\frac{\text{d}\tau_1}{\tau}$ and $\frac{\text{d}\tau_2}{\tau-\tau_1}$),
as well as a bath propagator momentum ${\bf q}$ with ${\abs{\bf q}} < k_F$ (probability: $\frac{\text{d}^3q}{(2k_F)^3}$). This fixes the time variable of the last piece to $\tau_3 = \tau-\tau_1-\tau_2$.
The whole process is illustrated in Fig.~\ref{fig:insert_mushroom}.
The Metropolis acceptance ratio $P_{\text{IM}}$ is
\begin{align}
 &\min \left(1, \frac{p_{\text{RM}}}{p_{\text{IM}}}  \frac{G^0_{\downarrow}(\tau_1,{\bf p})   \Gamma(\tau_2,{\bf p}+{\bf q})     G^0_{\downarrow}(\tau_3,{\bf p})   G^0_{\uparrow}(\tau_2,{\bf q})}
  { (2 \pi)^3  G^0_{\downarrow}(\tau,{\bf p}) \cdot \frac{1}{\tau} \frac{1}{\tau-\tau_1} \frac{1}{(2 k_F)^3} }\right).
\end{align}
The factor $(2\pi)^3$ in the denominator is part of the diagrammatic weight of the new configuration.
$p_{\text{IM}}$ and $p_{\text{RM}}$ are the probabilities of selecting the updates Insert-mushroom or Remove-mushroom, respectively.

\subsection{Remove-mushroom}
Remove-mushroom is the inverse update for Insert-mushroom. If the current diagram order is denoted by $N$, there are $N$ $T$ matrices which could be removed together with the corresponding impurity propagator.
However, the first $T$ matrix can not be removed, because it is never constructed by Insert-Mushroom. The same is true for the last $T$ matrix. That leaves $N-2$ possible $T$ matrices and balances the 
selection factors in the Metropolis algorithm.
Being the inverse update of Insert-Mushroom, the acceptance ratio of Remove-Mushroom is given by
\begin{align}
 &\min \left(1, \frac{p_{\text{IM}}}{p_{\text{RM}}}  \frac{ (2 \pi)^3  G^0_{\downarrow}(\tau,{\bf p}) \cdot \frac{1}{\tau} \frac{1}{\tau-\tau_1} \frac{1}{(2 k_F)^3} }
 {G^0_{\downarrow}(\tau_1,{\bf p})   \Gamma(\tau_2,{\bf p}+{\bf q})     G^0_{\downarrow}(\tau_3,{\bf p})   G^0_{\uparrow}(\tau_2,{\bf q})}
  \right).
\end{align}

\subsection{Reconnect}
\begin{figure}[tpb]
\includegraphics[width=0.99\linewidth]{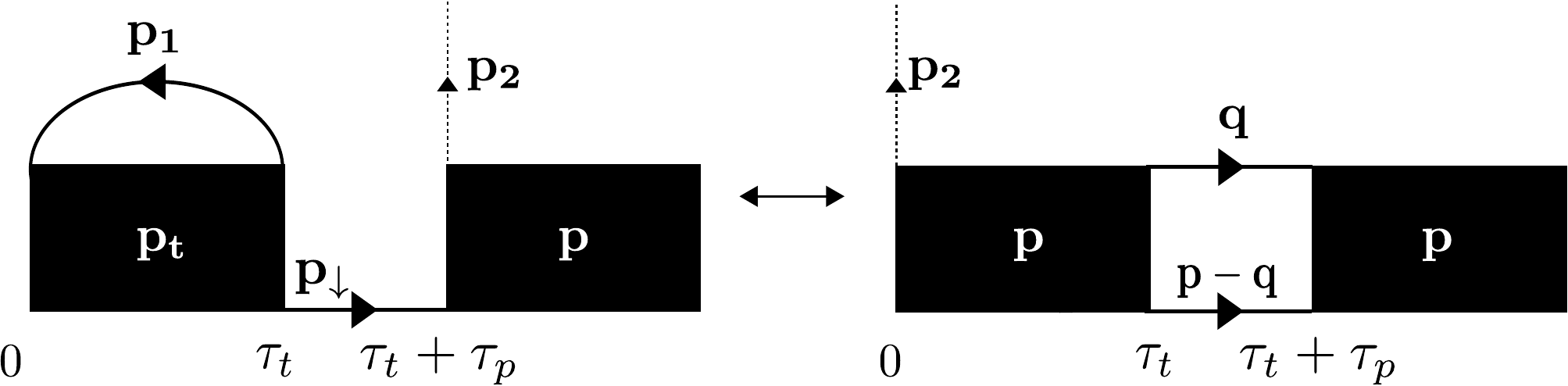}
\caption{\label{fig:reconnect1}  
Illustration of the first and second case of update Reconnect. The dotted vertical bath propagator line is connected to an arbitrary $T$ matrix in the diagram.}
\end{figure}
%
\begin{figure}[tpb]
\includegraphics[width=0.99\linewidth]{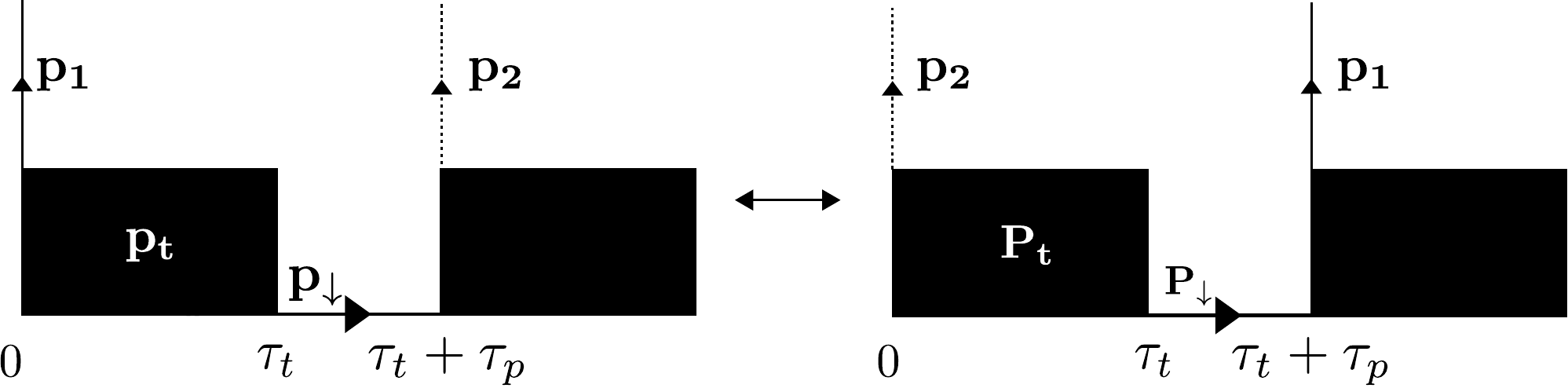}
\caption{\label{fig:reconnect2}  
Illustration of the third case of update Reconnect. The dotted and full vertical bath propagator lines are connected to arbitrary $T$ matrices in the diagram (as long as neither the full line of the left figure nor the 
dotted line of the right figure is connected to itself.}
\end{figure}

Reconnect is the key update of our procedure. 
It randomly selects one of the $T$ matrices, except the last one. In diagram order $N$, there are $N-1$ possible choices.
Suppose that a $T$ matrix with parameters $(\tau_t,{\bf p}_t)$ and an impurity neighbor adjacent to the right with parameters $(\tau_p,{\bf p}_\downarrow)$ is selected.
The update then proposes to swap the incoming bath propagator with the incoming bath propagator of its right neighbor.
There is a unique way of swapping as we are not working in cyclical representation.
Accordingly, the former bath propagator times $\tau_1$ and $\tau_2$ are exactly mapped on new times $\tau_1'$ and $\tau_2'$.
Index $1$ labels the bath propagator linked to the selected $T$ matrix.
Note that the mapping of the bath propagator momenta to the corresponding new momenta is not clear at this instant,
as the shape of the current topology has to be reflected.
In the moment of linking the new propagator configuration, the underlying momenta have to be adjusted in a manner described below.
Last, the resulting diagram has to be checked for one-particle-irreducibility -- the update has to be rejected if any impurity propagator line turns out uncovered.
Subsequent application of Reconnect updates allows to reach every bath propagator configuration and guarantees ergodicity.

More precisely, the update separates into three different cases depending on the current diagram configuration.
The first diagram topology (cf.~Fig.~\ref{fig:reconnect1}) is identified by having a mushroom-structure on the selected $T$ matrix -- its incoming bath line is connected with its
outgoing bath line. Since swapping will transfer a hole into a bath particle, it is necessary to create its new particle momentum ${\bf q}$.
This is done by uniform seeding on the interval $[-k_\text{max},k_\text{max}]$ for each component of $q$, where $k_\text{max}$ introduces the momentum cutoff of our
procedure. The update is rejected if $| {\bf q} | > k_\text{max}$ or if $|{\bf q}| < k_F$.
Concerning underlying momenta, the selected $T$ matrix is assigned the momentum ${\bf p}$ of the right neighboring $T$ matrix, while its right impurity
neighbor obtains ${\bf p}-{ \bf q}$. 
It is easy to compute final times
\begin{equation}
\begin{aligned}
\tau_1' &= \tau_2 - \tau_t - \tau_p \\ 
\tau_2' &= \tau_p .
\end{aligned}
\end{equation}
The acceptance ratio is
\begin{equation}
\label{eq::db1}
\min \left(1, \frac{   \Gamma(\tau_t,{\bf p})     G^0_{\downarrow}(\tau_p,{\bf p}-{\bf q})        G^0_{\uparrow}(\tau_1',{\bf p}_2)   G^0_{\uparrow}(\tau_2',{\bf q})  k_\text{max}^3   }
  {  \Gamma(\tau_t,{\bf p}_t)    G^0_{\downarrow}(\tau_p,{\bf p}_\downarrow)    G^0_{\uparrow}(\tau_1,{\bf p}_1)   G^0_{\uparrow}(\tau_2,{\bf p}_2)   k_F^3     } \right).
\end{equation}

The second topology is identified by a link between the outgoing end of the selected $T$ matrix and its right neighbor.
Being the inverse of the latter update, only one more step is necessary.
Instead of seeding new particle momentum, now the hole momentum has to be created on the selected $T$ matrix, thus explaining the factor of $k_F$ in Eq.~\ref{eq::db1}.
The acceptance ratio for the second topology is
\begin{align}
\min \left(1, \frac  {  \Gamma(\tau_t,{\bf p}_t)    G^0_{\downarrow}(\tau_p,{\bf p}_\downarrow)    G^0_{\uparrow}(\tau_1,{\bf p}_1)   G^0_{\uparrow}(\tau_2,{\bf p}_2)   k_F^3     }
{   \Gamma(\tau_t,{\bf p})     G^0_{\downarrow}(\tau_p,{\bf p}-{\bf q})        G^0_{\uparrow}(\tau_1',{\bf p}_2)   G^0_{\uparrow}(\tau_2',{\bf q})  k_\text{max}^3   } \right).
\end{align}

All other cases are included in the third topology (cf.~Fig.~\ref{fig:reconnect2}), defined by neither connecting the selected $T$ matrix with its right neighbor nor with itself.
Such cases are self-inverse.
No seeding is necessary, all bath momenta are purely swapped or added.
Determining final times is straightforward
\begin{equation}
\begin{aligned}
\tau_1' &= \tau_2 - \tau_t - \tau_p \\ 
\tau_2' &= \tau_1 + \tau_t + \tau_p .
\end{aligned}
\end{equation}
Note that holes are defined to have negative times.
Concerning momenta, only the selected $T$ matrix and its right impurity neighbor have to be considered, yielding new momenta ${\bf P}_t$ for $T$ matrix and
${\bf P}_\downarrow$ for impurity line:
\begin{equation}
\begin{aligned}
{\bf P}_t &= {\bf p}_t+{\bf p}_2-{\bf p}_1 \\ 
{\bf P}_\downarrow &= {\bf p}_\downarrow+{\bf p}_2-{\bf p}_1 \\
{\bf p}_1' &= {\bf p}_2 \\
{\bf p}_2' &= {\bf p}_1 .
\end{aligned}
\end{equation}
This results in the acceptance ratio
\begin{equation}
 \min \left(1,  \frac{   \Gamma(\tau_t,{\bf P}_t)     G^0_{\downarrow}(\tau_p,{\bf P}_\downarrow)        G^0_{\uparrow}(\tau_1',{\bf p}_2)   G^0_{\uparrow}(\tau_2',{\bf p}_1)   }
  {  \Gamma(\tau_t,{\bf p}_t)    G^0_{\downarrow}(\tau_p,{\bf p}_\downarrow)    G^0_{\uparrow}(\tau_1,{\bf p}_1)   G^0_{\uparrow}(\tau_2,{\bf p}_2)       } \right).
\end{equation}

\section{Partially bold diagrammatic Monte Carlo}
\label{sec:sec4}
In addition to the modifications of the bare code, we propose some changes only affecting the bold diagrammatic Monte Carlo routine.
Using full Green's functions as self-consistent input has the disadvantage of increasing sampling space drastically by 
forcing the tabulation of the full Green's function in both imaginary time and momentum. It is therefore beneficial to use the fact that first-order contributions dominate the fully bold propagator 
and construct a partially bold diagrammatic series out of quantities that are easily tabulated. 
It is possible to put the analytically known first-order self-energy into Dyson's equation in Matsubara frequency space to obtain the first-order Green's function.
A Fourier transform yields the basic propagator $G^1_\downarrow(\tau,{\bf p})$ without stochastic errors.
Only a few modifications are necessary to adjust the basic Monte Carlo routine:
First, every diagram containing at least one first-order self-energy diagram (cf.~Fig.~\ref{fig:fake}) has to be excluded from measurements. 
Second, it is no longer forbidden to connect adjacent $T$ matrices, just as in fully bold code\cite{prokofev2008B} -- the only difference is that the associated impurity weight 
is now $G^1_\downarrow - G^0_\downarrow$.

We also extended the molecule code to include partially bold propagators. This extension is almost identical to the bold polaron code.
Linking $T$ matrices is readmitted with the same (now non-artificial) weight $G^1_\downarrow - G^0_\downarrow$ for impurity propagators. The only new feature concerns the first-order molecule diagram that
is now sampled and has to be calculated with impurity weight $G^1_\downarrow - G^0_\downarrow$ . By this means, the ultraviolet divergence 
is cured and a second, independent molecule series is constructed which helps confirming robustness and reproducibility of the results.
Resummation is still needed for this new series. 

As a last comment on bold sampling, we would like to draw attention to the flaws of bold diagMC.
First, if the Dyson series is not absolutely convergent, the rearrangement of this series into the fully or partially bold series is potentially harmful, as
it constitutes a (though physically motivated) regrouping of terms which can yield any result for non-convergent series\cite{kozik2014}.
Second, many series in quantum field theory are asymptotic expansions, implying that results begin to get better and better with increasing 
expansion order until some maximum expansion order $N_{\text{max}}$ is reached, after which the factorial growth of the number of diagrams leads to huge oscillations.
In such a case, using a bare series is a common procedure, whereas the bold approach captures diagrams of higher order from the start.
Summarizing this line of argument, a bold diagrammatic approach seems only reasonable if the underlying series is convergent.

\section{Diagram regrouping and resummation}
\label{sec:sec5}
%
\begin{figure}[ptb]
\includegraphics[width=0.99\linewidth]{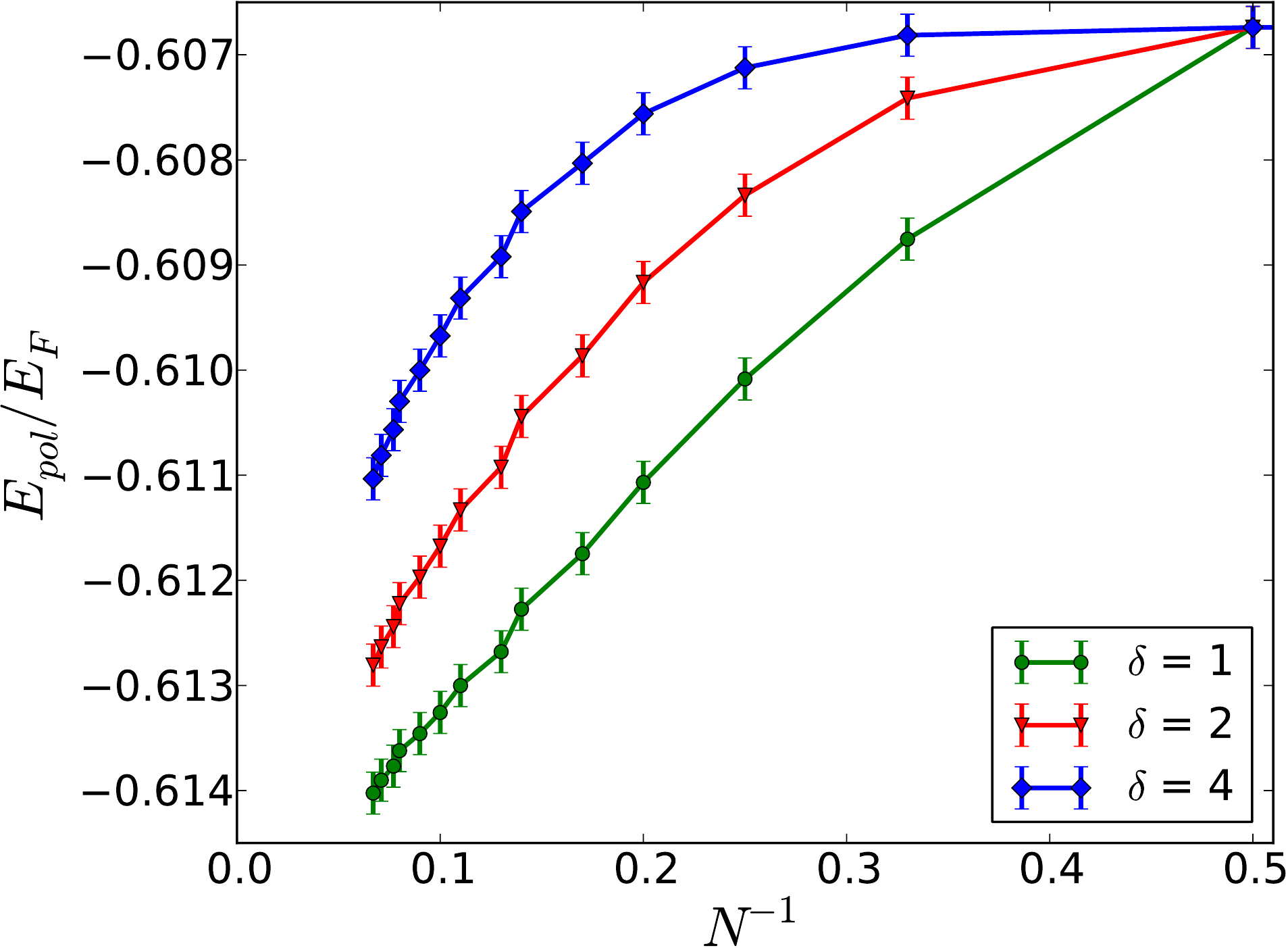}
\caption{\label{fig:resummationconventional} (Color online) 
Resummation methods of type Riesz with different exponents $\delta$ are compared for unitarity in a three-dimensional setup.
The plot shows polaron energies depending on maximum sampling order.}
\end{figure}
%
\begin{figure}[ptb]
\includegraphics[width=0.99\linewidth]{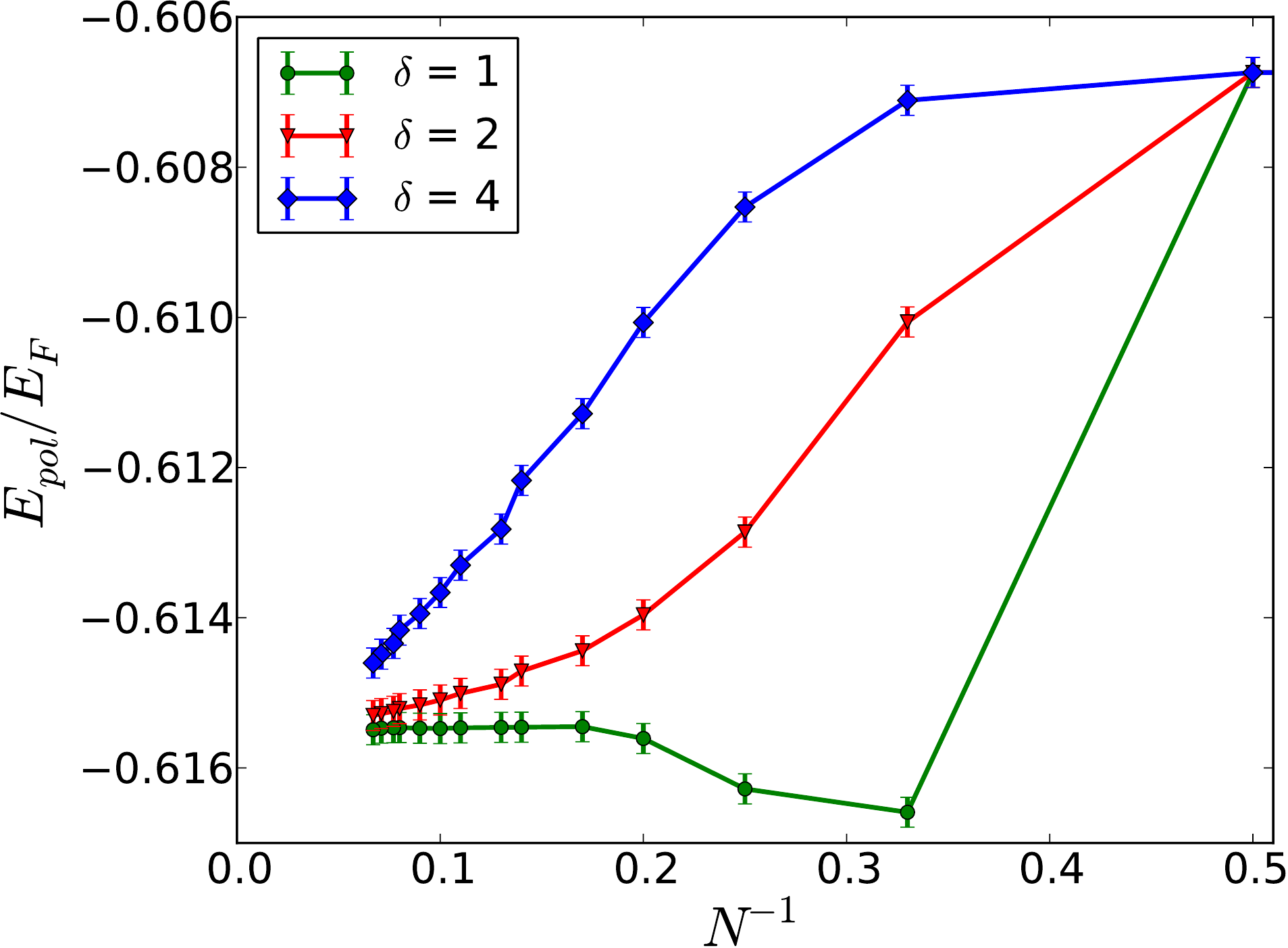}
\caption{\label{fig:resummationnewunitarity} (Color online) 
Resummation methods of type Riesz with different exponents $\delta$ applied to the modified bare series (see Eq.~\ref{eq:resum})
are compared at unitarity in a three-dimensional setup.
The plot shows polaron energies depending on maximum sampling order.}
\end{figure}

As diagrammatic expansions are in general not absolutely convergent, an important tool to study the underlying series is resummation\cite{prokofev2008B}.
This resummation procedure requires a discussion in more detail. Typically we find the molecular energies to be stable, but the polaron energies in the Bose-Einstein-condensate limit are harder.
Sharp resummations are potentially dangerous if the maximum sampling order is not high enough as can be seen as follows: With a strong resummation method, the produced curve is almost flat for low 
expansion orders and then bends down sharply for higher orders. Weak resummation methods on the other hand have more curvature for low expansion orders and flatten off if the order of divergence of 
the series is weak enough. There is thus a risk with strong resummation methods if only low expansion orders can be reached in the sense that a possibly strong curvature is missed, resulting in an 
apparently converging but wrongly extrapolated result in close vicinity to the first-order result. This effect is demonstrated\cite{energyestimator} in  Fig.~\ref{fig:resummationconventional} at unitarity.
Note that the bare series is monotonously decreasing which makes a high maximum resummation order necessary to extract the correct answer.
Summing it up: the more sign-alternating the bare series is, the better resummation works.

A typical resummation method is the Riesz resummation method.
These resummations will act upon self-energy series which can be written as $\mathcal{S} = \sum_N \mathcal{S}_{(N)}$, where $\mathcal{S}_{(N)}$ contains all contributions
of diagrams of order $N$. The order of a self-energy diagram is defined as the number of interaction $T$ matrices. 
The resummed self-energy series $\mathcal{S}'$ for some given maximum order $L$ is defined\cite{prokofev2008B} as 
\begin{equation}
\mathcal{S}'(L) = \sum_{N=1}^L \mathcal{S}_{(N)} F_N^{(L)},
\end{equation}
where $F$ is given by the Riesz coefficients
\begin{equation}
F_N^{(L)} = \left( \frac{L-N+1}{L} \right)^\delta.
\end{equation}
$\delta$ fixes the strength of the resummation: For $\delta = 0$, no resummation is performed at all, while only the first-order 
contribution is maintained in the limit $\delta \rightarrow + \infty$.
If this method is used on molecular series, it might be beneficial to set $F_2^{(L)} = 1$ as this ensures that the first contributing diagram (which is of second order for molecules)
always contributes with full weight.

We introduce a regrouping technique which seems to saturate much faster. Provided the series is absolutely convergent, this is always allowed. It is based on a regrouping of terms in the 
bare series in such a way that sign-alternation is maximized. 
The technique consists in splitting $\mathcal{S}_{(N)}$, the self-energy contributions of order $N$, into two parts:
\begin{equation}
\mathcal{S}_{(N)} = \mathcal{S}^r_{(N)} + \mathcal{S}^{ir}_{(N)},
\end{equation}
where $\mathcal{S}^r_{(N)}$ collects all diagrams containing at least one $T$ matrix linked by a hole to itself, cf.~Fig.~\ref{fig:fake},
and $\mathcal{S}^{ir}_{(N)}$ gathers the rest.
We propose a new series $\mathcal{S}' = \sum_N \mathcal{S'}_{(N)}$ which aims to maximize sign-alternation in $\mathcal{S'}_{(N)}$.
The coefficients in the resummation procedure depend on the expansion order for $k \in \mathbb{N}$ as
\begin{equation}
\begin{aligned}
\mathcal{S'}_{(1)} &= \mathcal{S}_{(1)}, \\
\mathcal{S'}_{(N = 2k+1)} &= \mathcal{S}^{ir}_{(N)} + \frac{1}{2} \mathcal{S}^{ir}_{(N-1)} +  \frac{1}{2} \mathcal{S}^r_{(N)}, \\
\mathcal{S'}_{(N = 2k)} &= \mathcal{S}^{r}_{(N)} +  \frac{1}{2}  \mathcal{S}^{r}_{(N-1)} +  \frac{1}{2} \mathcal{S}^{ir}_{(N)}. \label{eq:resum}
\end{aligned}
\end{equation}
These coefficients are in principle arbitrary -- our choice was designed to show fast saturation as can be seen in Fig.~\ref{fig:resummationnewunitarity} for the case of the Fermi-polaron at unitarity: 
The application of conventional Riesz resummations on the reordered series illustrates
that the manually implemented oscillations make it possible to read off polaron energies reliably and allow for a clear statement whether the expansion order is high enough or not.
When reverting the roles of reducible and irreducible in Eq.~\ref{eq:resum}, the same answer can be found but only after a stronger resummation. 
Note that although the sum of diagrams of a specific order of the unitary polaron series is vanishing within error bars, the different terms are not small.

As the regrouped series agrees with the results of the standard bare series, this might indicate that the polaron Dyson series is convergent or that the maximum expansion order $N_{\text{max}}$ of its asymptotic
expansion is essentially infinity at unitarity. Nevertheless, Fig.~7 of Ref.~\onlinecite{vlietinck2013} suggests that the series might be asymptotic in fact, as a doubly bold regrouping shows clear signs 
of growing fluctuations for increasing expansion order after an initial improvement of results.

\section{Results}
\label{sec:sec6}
In this section, our diagrammatic Monte Carlo results for a mass-imbalanced polaron are presented. 

\subsection{Polaron energy and residue at unitarity}

\begin{figure}[ptb]
\includegraphics[width=0.99\linewidth]{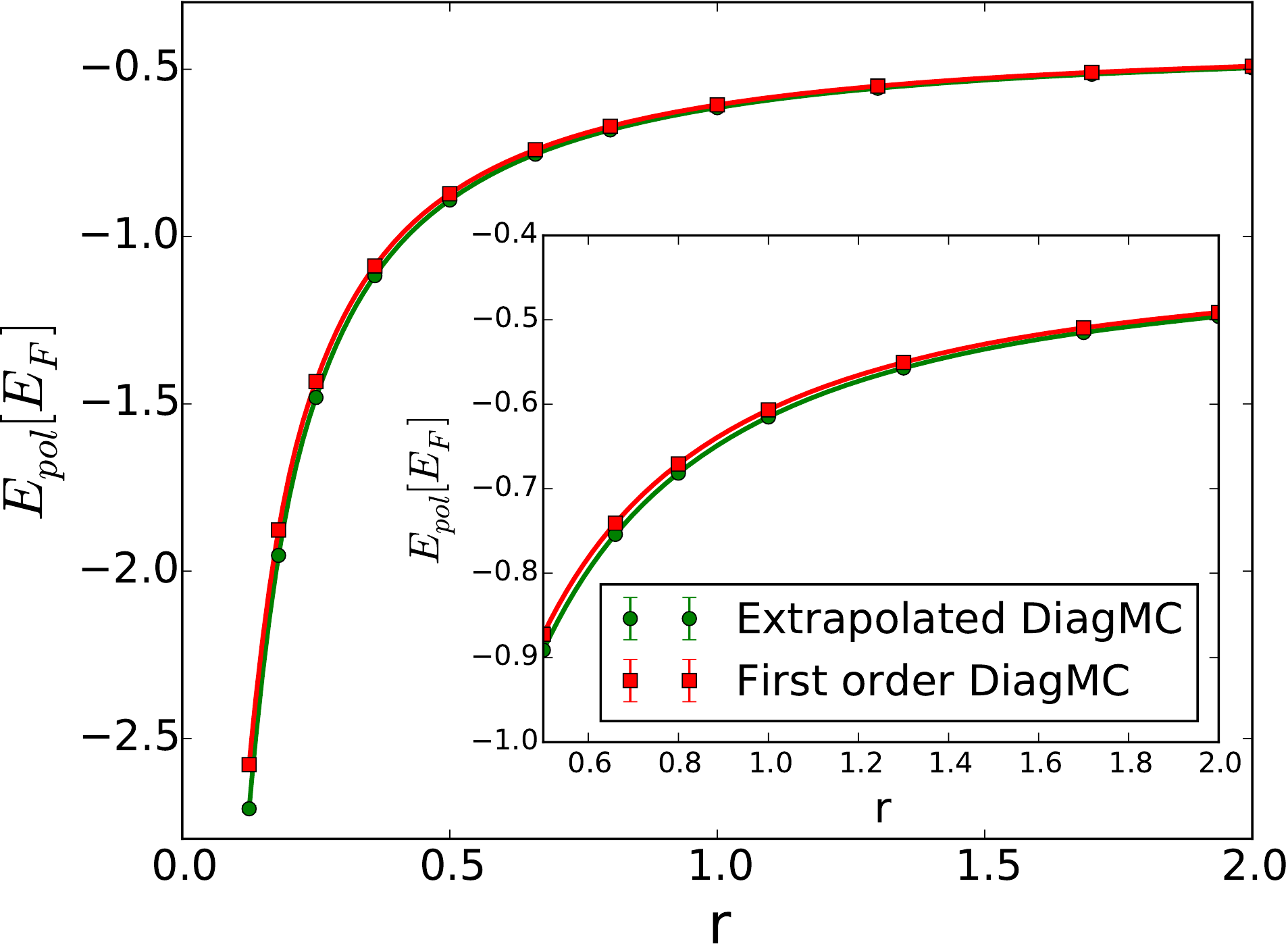}
\caption{\label{fig:energy} (Color online)  
Polaron energy at unitarity for different mass ratios $r$. The inset shows the flat part of the figure. Many-body effects get more pronounced for a lighter polaron.}
\end{figure}
%
\begin{figure}[ptb]
\includegraphics[width=0.99\linewidth]{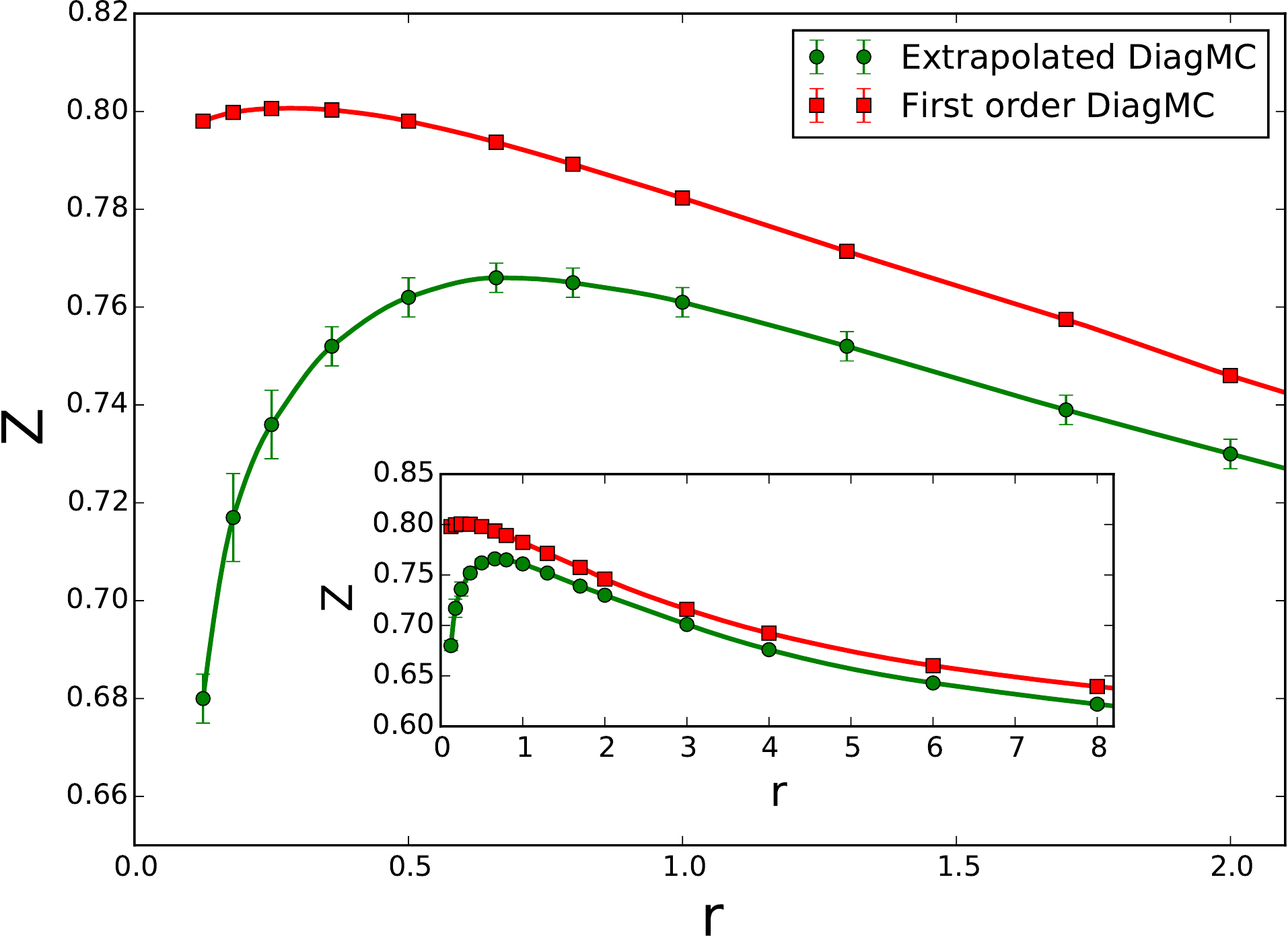}
\caption{\label{fig:residue} (Color online)  
Polaron residue at unitarity for different mass ratios $r$. The whole range of masses shows a clear difference between the first order result and the full diagrammatic answer. 
The inset shows the curve for additional values of $r$.}
\end{figure}

Fig.~\ref{fig:energy} plots the polaron energy at unitarity for different mass ratios $r = \frac{m_\downarrow}{m_\uparrow}$. While the variational Chevy ansatz, here labeled as
first order, captures the whole curve qualitatively, its quantitative accuracy gets less precise for low $r$, i.e., a light polaron.
Note that this energy curve reproduces the correct infinite mass limit\cite{combescot2007} $E_{\text{pol}} = -0.5$ for an imbalance ratio as low as $r = 2$.
For the case of an immobile impurity, the polaron is subject to Anderson's orthogonality catastrophe\cite{anderson1967} and the quasiparticle description is no longer appropriate.
For a light impurity, the polaron energy decreases rapidly as the effective interaction is stronger for smaller reduced mass (see Appendix \ref{sec:appA}) at unitarity. The next subsection 
will show that this effect is weakened for finite scattering length.
Eventually, a very light impurity will be subject to relativistic effects so that our description will no longer be appropriate.
The extraction of error bars was based on conservative extrapolations of light Riesz resummation with $\delta = 1$. For details, consult appendix \ref{sec:appB}.

The quasiparticle residue $Z$ can be extracted from the asymptotic decay of the full propagator\cite{prokofev2008B} 
$G_\downarrow(\tau,{\bf k}) \underset{\tau \rightarrow \infty}{\longrightarrow} - Z e^{-(E-\mu_\downarrow) \tau}$,
where $E$ is the ground state energy of the quasiparticle. This form implies that a suitable Monte Carlo estimator for the residue is $Z = (1 + A(E, {\bf p}))^{-1}$ with 
\begin{equation}
A(E, {\bf p}) = - \int_0^\infty \tau \mathcal{S}(\tau, {\bf p}) e^{(E-\mu_\downarrow)  \tau} \text{d}\tau.
\end{equation}
$\mathcal{S}$ is the sampled self-energy.

Our results for $Z$ are depicted in Fig.~\ref{fig:residue}.
In this case, the first-order ansatz is both quantitatively and qualitatively different, predicting a different position of
the mass-imbalance ratio of maximum residue. This might indicate that the variational wave-function description works particularly well for energy based quantities,
while it might take further particle-hole terms to capture the residue equally well.
Therefore the quasiparticle with maximum residue can be found at higher $r$ than estimated by first order. 
At low $r$, higher orders affect $Z$ stronger and stronger, down to a ratio as low as $r = 0.125$ 
which is sufficient for most mixtures, e.g., ~$^6$Li--$^{40}$K. In this regime, the differences between the diagrammatic answer and the first order result are most pronounced.
For high $r$, the diagMC solution yields a roughly constant shift to the first order answer.

As a trimer state acquires more and more strength with respect to the polaron state for decreasing $r$ and as the molecular state is strengthened for increasing\cite{parish2011} $r$, it seems
natural that the polaron residue takes on a maximum value in between -- once it is no longer the ground state, its residue will decay quickly (although it will not be zero because there is no
simple decay channel\cite{schmidt2011}).
The measured residue is strictly lower than the first order variational result. This is remarkable as the Functional Renormalization Group analysis of Ref.~\onlinecite{schmidt2011} 
predicts a higher residue for unitarity at $r = 1$. Further investigation is needed to understand this discrepancy.

No resummation was used to extract the quasiparticle residues. For $r \gtrsim 0.5$, the series seemed to saturate within our maximum expansion order.
The extrapolation error was approximated to be twice the fluctuation of the saturating points.
For $r \lesssim 0.5$, the series changed and the saturation was not visible anymore. 
These points are therefore only valid if a linear extrapolation to 
infinite expansion order is appropriate. This extrapolation error was approximated by the method explained in appendix \ref{sec:appB}.

\subsection{Polaron energy beyond unitarity}
\begin{figure}[ptb]
\includegraphics[width=0.99\linewidth]{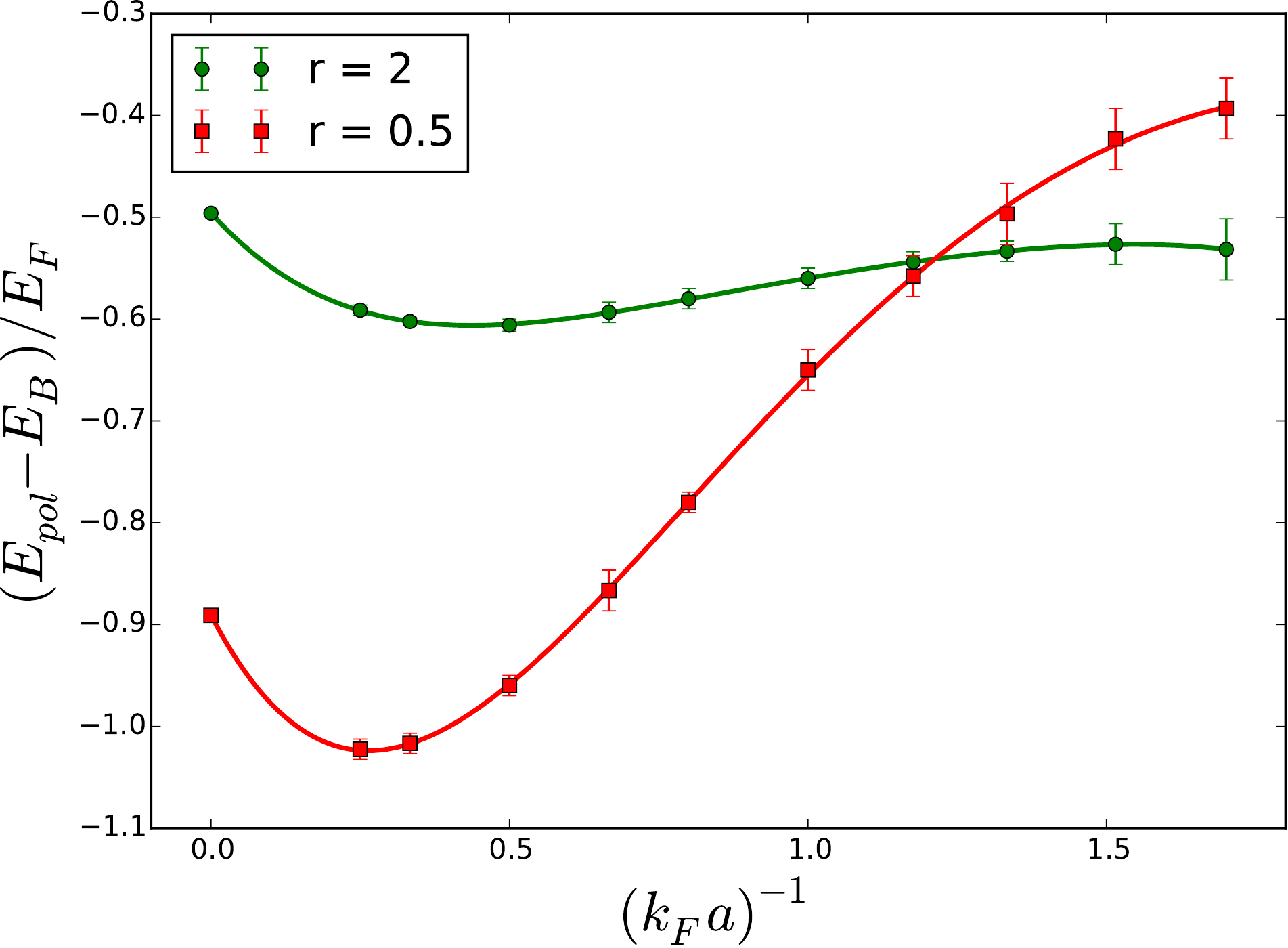}
\caption{\label{fig:energy_coupling} (Color online) The polaron energy of two values of $r$ is plotted for various interaction parameters $k_F a$. Note that the binding energy $E_B = (2 m_r a^2)^{-1}$
is subtracted. For strong interactions, the light impurity acquires a higher effective dressing relative to the binding energy than the heavy impurity. This is eventually reversed in the BEC-regime.
}
\end{figure}
%
Fig.~\ref{fig:energy_coupling} shows polaron energies at varying coupling strength in the BEC-regime for two different mass-imbalance ratios $r$.
Both curves experience a peak of maximum dressing around $(k_F a)^{-1} = 0.4$. Decreasing the coupling further, this relative energy is decreased for both masses,
although the light impurity is affected more strongly. Eventually, the heavy impurity has a higher effective dressing (relative to the binding energy) than the light impurity.
This is a consequence of the $m_r/(2\pi a)$ term in the denominator of the $T$ matrix (see Appendix \ref{sec:appA}) that strengthens the effective interaction between impurity and bath atoms for increasing $r$
at a given interaction strength. Note that these curves extend into the molecular sector\cite{parish2011} where the polaron ceases to be the ground state.
Concerning the residue, no qualitative difference could be seen between the $r=0.5$ and $r=2$ curves.

\subsection{Two-particle-hole channel}
\begin{figure}[ptb]
\includegraphics[width=0.99\linewidth]{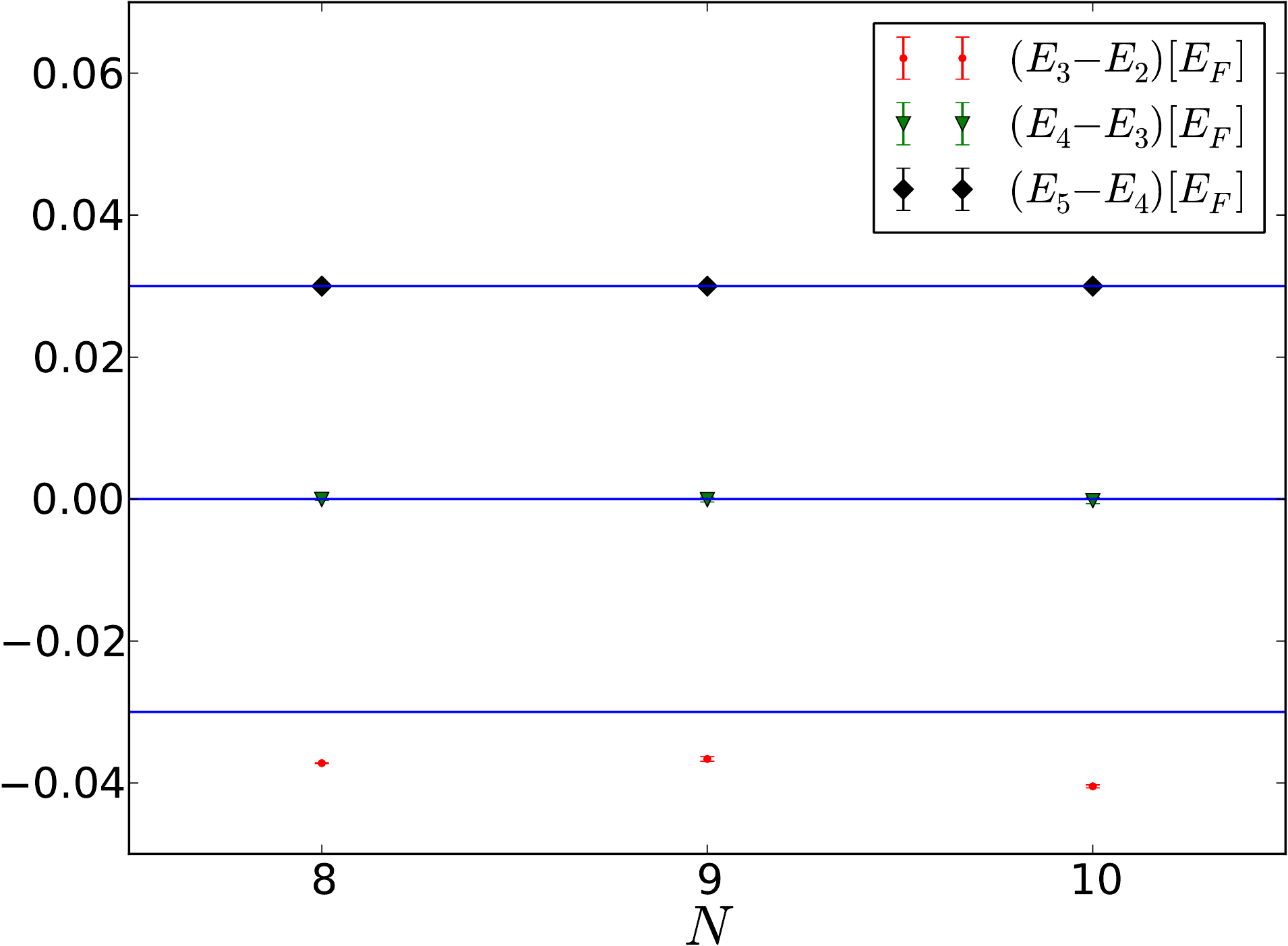}
\caption{\label{fig:hole_plot} (Color online)  
Different contributions to the full polaron energy are compared for three maximum expansion orders $N$. $E_n$ denotes contributions including up to n-particle-hole diagrams.
Two of the curves have been offset by $\pm 0.03 E_F$ for clarity. The points were measured for a mass-imbalance ratio $r=0.125$ at unitarity.}
\end{figure}
%
For quasi-two-dimensional geometries, a remarkable quantitative precision of two-particle-hole wave-functions was found for polaron energies\cite{kroiss2014,parish2013}.
A natural question is whether this approach remains valid in three dimensions. Fig.~\ref{fig:hole_plot} compares different particle-hole channels for three maximum expansion orders.
For the selected measurement point (unitarity with imbalance $r = 0.125$), the three-particle-hole channel contributes with slight quantitative differences, 
whereas four-particle-hole and five-particle-hole diagrams vanish within error bars.
This confirms the observation that the two-particle-hole result is essentially correct and can be used for quantitative calculations.
The classification of diagrams into particle-hole classes breaks down for the fully bold approach, because each bold diagram captures bare diagrams of different particle-hole order.
For the partially bold scheme, this does not apply since the diagrammatic structure of the partially bold Green's functions ensures that the propagation will start with zero holes and is guaranteed 
to switch to the 1-ph sector at least once.

\subsection{Spectral function}
%
\begin{figure}[ptb]
\includegraphics[width=0.99\linewidth]{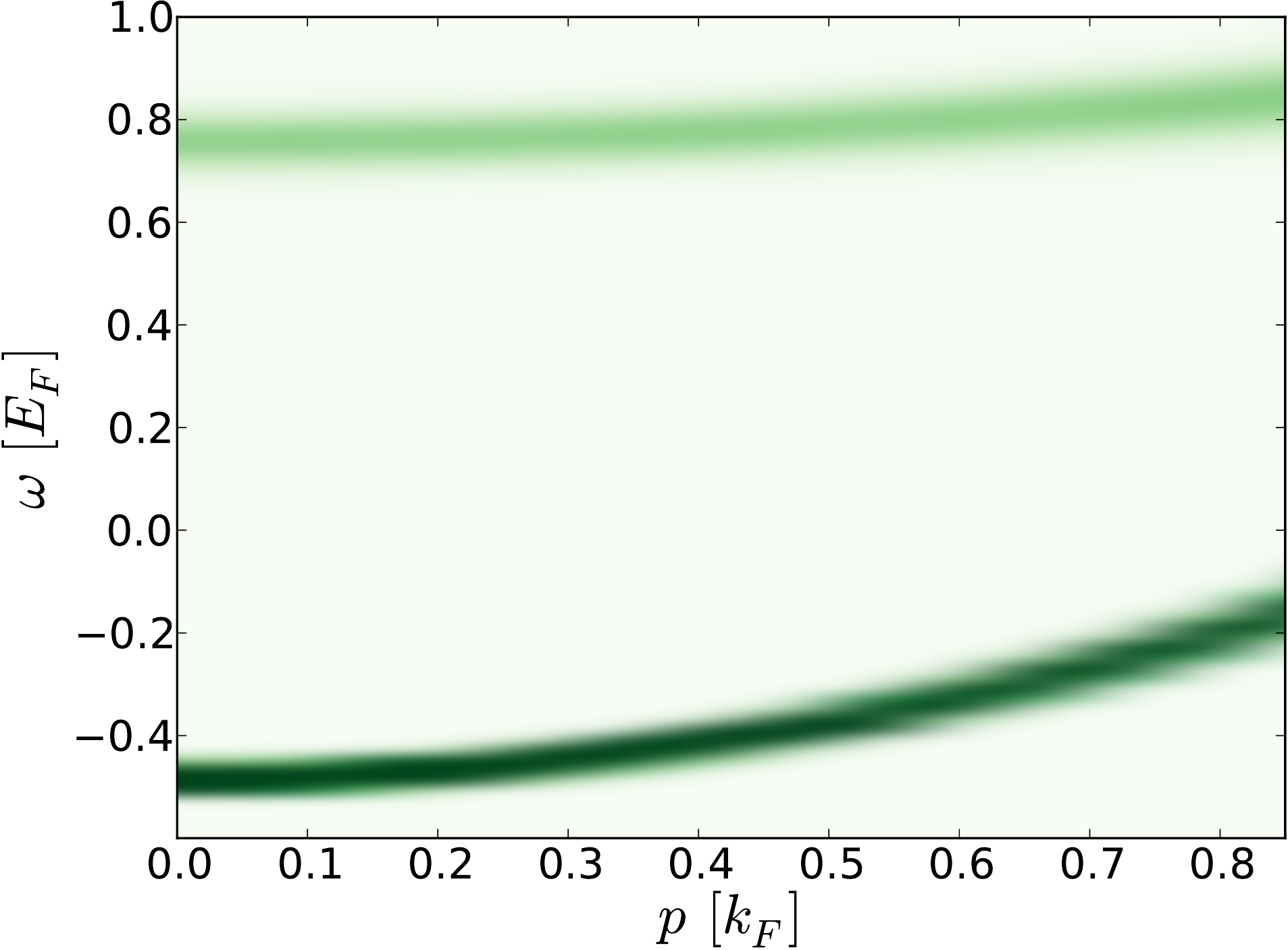}
\caption{\label{fig:spectral_function} (Color online)  
The polaron spectral function is plotted for a mass-imbalance $r = 2$ at unitarity. Note the quadratic dispersion and the repulsive polaron at positive energies.}
\end{figure}
%
It is possible to extract the polaronic spectral function $\mathcal{A}(\omega,{\bf p}) = -2 \ \text{Im}{\left( \omega + i0 - \epsilon_{{\bf p}} +  \mu_\downarrow^0  - \mathcal{S}(\omega,{\bf p}) \right)^{-1}}$ 
from the sampled self-energy $\mathcal{S}$. The full Green's function can be calculated by Fourier transform, subsequent application of Dyson's equation and another Fourier transform.
Then, analytic continuation to real quantities is applied\cite{gubernatis1996,alps2}.
Fig.~\ref{fig:spectral_function} presents the spectral function for a mass-imbalance ratio $r = 2$. Although the energy corresponds to the infinite mass limit\cite{combescot2007}, the polaron 
remains a stable quasiparticle. The dispersion follows a parabola with positive effective mass, whereas at higher energies, 
the repulsive polaron (a metastable eigenstate of the Hamiltonian) can clearly be seen\cite{schmidt2011,massignan2011}.

\subsection{Quantitative exactness of variational energies}
%
\begin{figure}[ptb]
\includegraphics[width=0.99\linewidth]{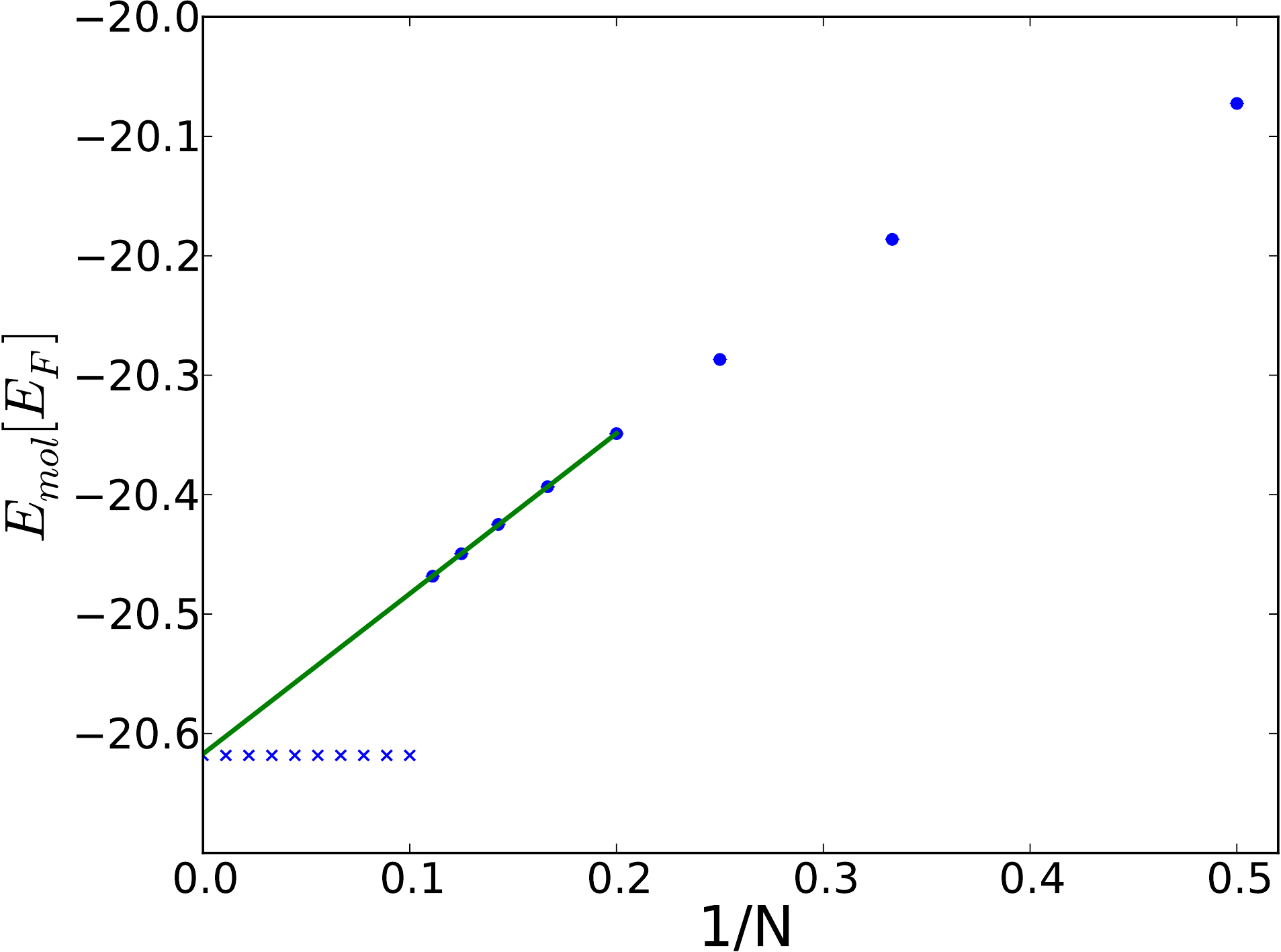}
\caption{\label{fig:massimbmol} (Color online)  
Molecule energy extrapolation for $k_F a = 0.5$ and $r = 0.25$. The maximum expansion order is denoted by $N$. The solid line corresponds to a linear fit of the last five points.
The crosses mark the one-particle-hole result of Ref.~\onlinecite{parish2011}. Riesz resummation with exponent $\delta = 6$ was used.}
\end{figure}
%
\begin{figure}[ptb]
\includegraphics[width=0.99\linewidth]{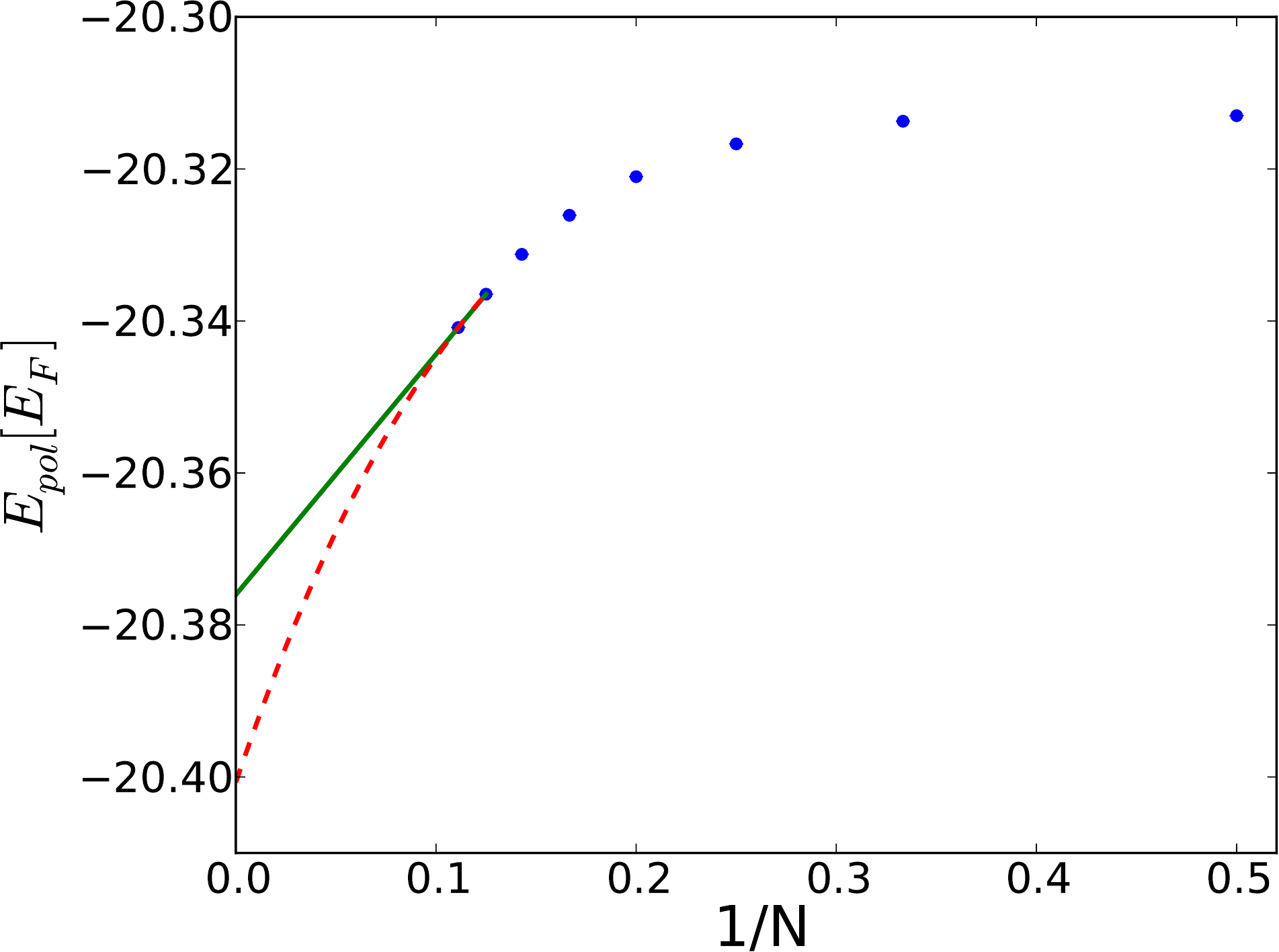}
\caption{\label{fig:massimbpol} (Color online)  
Polaron energy extrapolation for $k_F a = 0.5$ and $r = 0.25$. The maximum expansion order is denoted by $N$. The solid line corresponds to a linear fit of the last five points, 
the dashed line is a special fitting function explained in Appendix \ref{sec:appB}. The one-particle-hole result of Ref.~\onlinecite{parish2011} is identically with the point $N = 2$.
Riesz resummation with exponent $\delta = 4$ was used.}
\end{figure}
%
In this subsection, the extraction of polaron and molecule energies by resummation is compared to variational one-particle-hole wave-functions.
We choose the point $k_F a = 0.5$ and $r = 0.25$ of the mass-imbalanced phase diagram\cite{parish2011} as it stays away from the peculiarities of unitarity and the trimer
threshold. 
Molecular energies (Fig.~\ref{fig:massimbmol}) yield perfect agreement with the variational ansatz and motivate
the quantitative correctness of variational wave-functions for the Fermi-polaron problem.
For the polaron (Fig.~\ref{fig:massimbpol}), using a one-particle-hole wave-function underestimates its energy slightly.
Hence, it would be beneficial to use at least two-particle-hole precision for the polaron sector for a precise mapping of the phase diagram.
Altogether, the phase diagram of Ref.~\onlinecite{parish2011} can be expected to be nearly quantitatively exact. Nevertheless, as the polaron phase is underestimated, it will be shifted
into the molecular sector. As this will reduce the small size of the nonzero-momentum molecular phase (labeled as Fulde-Ferrell-Larkin-Ovchinnikov phase, FFLO) further, it remains open whether this phase really exists.
We suggest to compute the phase boundary with high precision within a variational 2-ph polaron approach.

\subsection{Dimensionless contact coefficient}
Tan's contact coefficient\cite{punk2009} $\mathcal{C}$ for a
strongly population-imbalanced Fermi gas is linked to the
dimensionless contact coefficient $s$ by
\begin{equation}
\mathcal{C}=s\cdot k_{F,\downarrow}^3 k_{F,\uparrow}.
\end{equation}
Here, $k_{F,\downarrow}$ is the Fermi momentum of the minority species which is finite for the strongly imbalanced Fermi gas.
The dimensionless contact coefficient $s$ can be accessed easily by calculating the derivative of the polaron energy with respect to the
dimensionless coupling\cite{punk2009} $(k_{F,\uparrow}a)^{-1}$.
The resulting contact curve agrees with first order calculations within error bars.

\section{Conclusion}
\label{sec:sec7}
Our work extends the diagrammatic Monte Carlo polaron routines to the more general case of a mass-imbalanced polaron.
This is an important limiting case of population imbalanced Fermi gases and allows to estimate key properties of its phase diagram. 
The diagrammatic space can be drastically reduced by sampling the self-energy instead
of the full Green's functions and by using the ladder approximation as basic interaction element.
We presented a critical analysis and alternative check of diagrammatic Monte Carlo
as well as a partially bold approach, thus broadening the toolbox of the diagrammatic Monte Carlo method.
An additional regrouping technique was presented to speed up extrapolation to infinite diagram order for absolutely convergent series.
While the first-order variational ansatz could give qualitative and quantitative good results for the polaron
energy at different polaron masses, discrepancies are more pronounced for the polaron residue.
For this quantity, higher orders have to be included in order to capture the whole physics.
Concerning Tan's contact coefficient, an excellent agreement was found with the Chevy variational wave-function.
The polaronic spectral function was extracted from imaginary time representation of diagrammatic Monte 
Carlo data by means of analytic continuation. It demonstrates a clean parabolic dispersion as well as the existence of the repulsive polaron. 
The two-particle-hole wave-function ansatz, which was shown to be essentially exact in quasi-two-dimensional geometries\cite{kroiss2014},
provides an equally good description of quasi-particle energies in three dimensions. Therefore, using one-particle-hole trial 
functions will lead to a phase diagram which overestimates molecular contributions and might lead to a weakening of the FFLO state found in Ref.~\onlinecite{parish2011}.

We are grateful to M. Bauer, T. Enss, J. Levinsen, M. Parish, N. Prokof'ev, R. Schmidt, B. Svistunov,
K. Van Houcke and W. Zwerger for valuable discussions. We would also like to thank M. Parish for sharing her data with us.
This work is supported by the Excellence Cluster NIM, FP7/Marie-Curie Grant No. 321918 (FDIAGMC), FP7/ERC Starting Grant No.
306897 (QUSIMGAS) and by a grant from the Army Research Office with funding from DARPA.
\appendix
\section{Many-body $T$ matrix for unequal masses}
\label{sec:appA}
It is possible to write the $T$ matrix in presence of the Fermi sea in a self-consistent way\cite{kroiss2012}
\begin{equation}
\begin{aligned}
 &- i \Gamma(\omega,{\bf p}) = - i g \\
 &+ \int_{q > k_F} \frac{d^3q}{(2\pi)^3} \int \frac{dq^0}{2\pi} 
 (-i g ) \\
 &i G^0_{\uparrow}(q^0,{\bf q}) i G^0_{\downarrow}(\omega-q^0,{\bf p}-{\bf q}) (-i \Gamma(\omega,{\bf p})).
\end{aligned}
\end{equation}
Since \(T\) is replacing an interaction line in first approximation it has to follow the same sign convention as \(g\).
Also note that for a point interaction, \(g({\bf k})\) is constant and independent of momentum. Rewriting yields
\begin{equation}
\begin{aligned}
&\frac{1}{\Gamma(\omega,{\bf p})} = \\
&\frac{1}{g} - i \int_{q > k_F} \frac{d^3q}{(2\pi)^3} \int \frac{dq^0}{2\pi}
G^0_{\downarrow}(\omega-q^0,{\bf p}-{\bf q}) G^0_{\uparrow}(q^0,{\bf q}).
\end{aligned}
\end{equation}
The integral is calculated by residue calculus; as only the pole of \(G^0_{\downarrow}\) is in the upper plane, it follows
\begin{equation}
\begin{aligned}
\label{tmatrixdiag}
&\frac{1}{\Gamma(\omega,{\bf p})} = \\
&\frac{1}{g} - \int_{q > k_F} \frac{d^3q}{(2\pi)^3} \frac{1}{\frac{q^2}{2m_{\uparrow}}+\frac{({\bf p}-{\bf q})^2}{2m_\downarrow}-E_F-\mu_\downarrow^0-\omega}.
\end{aligned}
\end{equation}
The problem of the latter expression is an ultraviolet divergence in \(q\), consequently it has to be regularized.

A similar series appears in the standard scattering Lippmann-Schwinger equation and can be used for regularization.
Starting point is the Lippmann-Schwinger form of the relative motion Schr\"odinger equation
\begin{equation}
 \ket{\psi_{{\bf k}}} = \ket{{\bf k}} + \frac{1}{E-\hat{H}_0+i0} \hat{V} \ket{\psi_{{\bf k}}}.
\end{equation}
Imposing \(\hat{V}\) on the left and 
introducing the two-body $T$ matrix \(\hat{T}^{2B}\) via \(\hat{T}^{2B}\ket{{\bf k}} \equiv \hat{V} \ket{\psi_{{\bf k}}}\), it follows
\begin{equation}
 \hat{T}^{2B} \ket{{\bf k}} = \hat{V} \ket{{\bf k}} + \hat{V} \frac{1}{E-\hat{H}_0+i0} \hat{T}^{2B}\ket{{\bf k}}.
\end{equation}
In the next step, \(\bra{{\bf k'}}\) is multiplied from the left with \( \abs{{\bf k}} = \abs{{\bf k'}}\) and a complete set of eigenfunctions of \(\hat{H}_0\) is inserted
\begin{equation}
\begin{aligned}
 &\bra{{\bf k'}}\hat{T}^{2B} \ket{{\bf k}} =  \bra{{\bf k'}} \hat{V} \ket{{\bf k}} \\
 &+ \int \frac{d^3h}{(2\pi)^3}  \bra{{\bf k'}}\hat{V} \ket{{\bf h}} \bra{{\bf h}} \frac{1}{E-\hat{H}_0+i0} \hat{T}^{2B}\ket{{\bf k}} .
 \end{aligned}
\end{equation}
The matrix elements of the potential are trivial for a pseudo-potential and the newly inserted \(\bra{{\bf h}}\) is an eigenstate of 
\(\hat{H}_0\)
\begin{equation}
 \bra{{\bf k'}}\hat{T}^{2B} \ket{{\bf k}} = g +
 g \int \frac{d^3h}{(2\pi)^3}  \frac{1}{E-\frac{h^2}{2 m_r}+i0} \bra{{\bf h}} \hat{T}^{2B}\ket{{\bf k}}.
\end{equation}
Looking at \(\bra{{\bf h}} \hat{T}^{2B}\ket{{\bf k}} \) more closely
\begin{equation}
\begin{aligned}
 &\bra{{\bf h}} \hat{T}^{2B}\ket{{\bf k}} = \melement{{\bf h}} {\hat{V}} {\psi_{{\bf k}}} \\
 &= \int \frac{d^3k''}{(2\pi)^3} \bra{{\bf h}} \hat{V}\ket{{\bf k''}}\braket{{\bf k''}}{\psi_{{\bf k}}} 
 = \int \frac{d^3k''}{(2\pi)^3}  g \braket{{\bf k''}}{\psi_{{\bf k}}} \\
 &= \int \frac{d^3k''}{(2\pi)^3} \bra{{\bf k'}} \hat{V}\ket{{\bf k''}}\braket{{\bf k''}}{\psi_{{\bf k}}} 
 = \bra{{\bf k'}} \hat{T}^{2B}\ket{{\bf k}}.
\end{aligned}
\end{equation}
Remarkably, this expression does not depend on the first element.
As there is no \({\bf h}\)-dependence left, the $T$ matrix can be extracted from the integral
\begin{equation}
\label{fT}
 \frac{1}{\bra{{\bf k'}} \hat{T}^{2B}\ket{{\bf k}}} = \frac{1}{g} - \int \frac{d^3h}{(2\pi)^3} \frac{1}{E-\frac{h^2}{2 m_r}}.
\end{equation}
The two-body $T$ matrix is related to the scattering length \cite{stoof2009} if the effective range can be set to zero
\begin{equation}
 \frac{1}{\bra{{\bf k'}} \hat{T}^{2B}\ket{{\bf k}}} = \frac{m_r}{2 \pi a} (1 + i a k).
\end{equation}
\(k\) is related to the energy of relative motion and can be generalized to off-shell behavior
\begin{equation}
 \frac{k^2}{2 m_r} = E = E_{\text{tot}} - E_{\text{com}} = \omega + E_F + \mu^0_{\downarrow} - \frac{p^2}{2 M}.
\end{equation}
Note that the total energy \(\omega\) is measured with respect to \(E_F\) and \(\mu^0_{\downarrow}\). \(p\) is the total momentum of the
system. Inserting this into equation \ref{fT} yields
\begin{equation}
\begin{aligned}
\label{tmatrixlipp}
  \frac{1}{g} &= \frac{m_r}{2\pi a} -\frac{m_r}{2\pi} \sqrt{\frac{m_r}{M}p^2-2 m_r (\omega+E_F+\mu^0_{\downarrow})} \\
 &+ \int \frac{d^3h}{(2\pi)^3} \frac{1}{\frac{p^2}{2 M}+\frac{h^2}{2 m_r} - \omega - E_F - \mu^0_{\downarrow}}.
\end{aligned}
\end{equation}

Finally, a suitable expression for the $T$ matrix can be found by inserting equation \ref{tmatrixlipp} into equation \ref{tmatrixdiag}
\begin{align}
 &\Gamma(\omega,{\bf k})^{-1} = \nonumber \\
 &\frac{m_{r}}{2 \pi a} - \frac{m_r}{2 \pi} \sqrt{\frac{m_r}{M} k^2 - 
 2 m_r (E_F + \mu_{\downarrow}^{0} + \omega)} \nonumber  \\
 &+ \int  \frac{d^3 q}{(2 \pi)^3} \frac{1}{\frac{k^2}{2 M} + \frac{({\bf q}-\frac{{\bf k}}{2})^2 }{2 m_r} - E_F -\mu_{\downarrow}^{0} -
 \omega} \nonumber \\
  &- \int_{q>k_F} \frac{d^3 q}{(2 \pi)^3} \frac{1}{\frac{q^2}{2 m_{\uparrow}} 
  + \frac{({\bf k}-{\bf q})^2 }{2 m_{\downarrow}} - E_F -\mu_{\downarrow}^{0} -  \omega}.
\end{align} 
This can be cast into its final form by a shift ${\bf q} \rightarrow {\bf q} + \frac{{\bf k}}{2} - \frac{m_\uparrow{\bf k}}{M}$ in the first integral
\begin{align} 
 & \Gamma(\omega,{\bf k})^{-1} = \nonumber \\
 &\frac{m_{r}}{2 \pi a} - \frac{m_r}{2 \pi} \sqrt{\frac{m_r}{M} k^2 - 
 2 m_r (E_F + \mu_{\downarrow}^{0} + \omega)} \nonumber  \\
 &+ \int_{q<k_F}  \frac{d^3 q}{(2 \pi)^3} \frac{1}{\frac{q^2}{2 m_{\uparrow}} + \frac{({\bf k}-{\bf q})^2 }{2 m_{\downarrow}} - E_F -\mu_{\downarrow}^{0} -  \omega}.
\end{align}

\section{Extrapolation of resummed data}
\label{sec:appB}
%
\begin{figure}[tbp]
\includegraphics[width=0.99\linewidth]{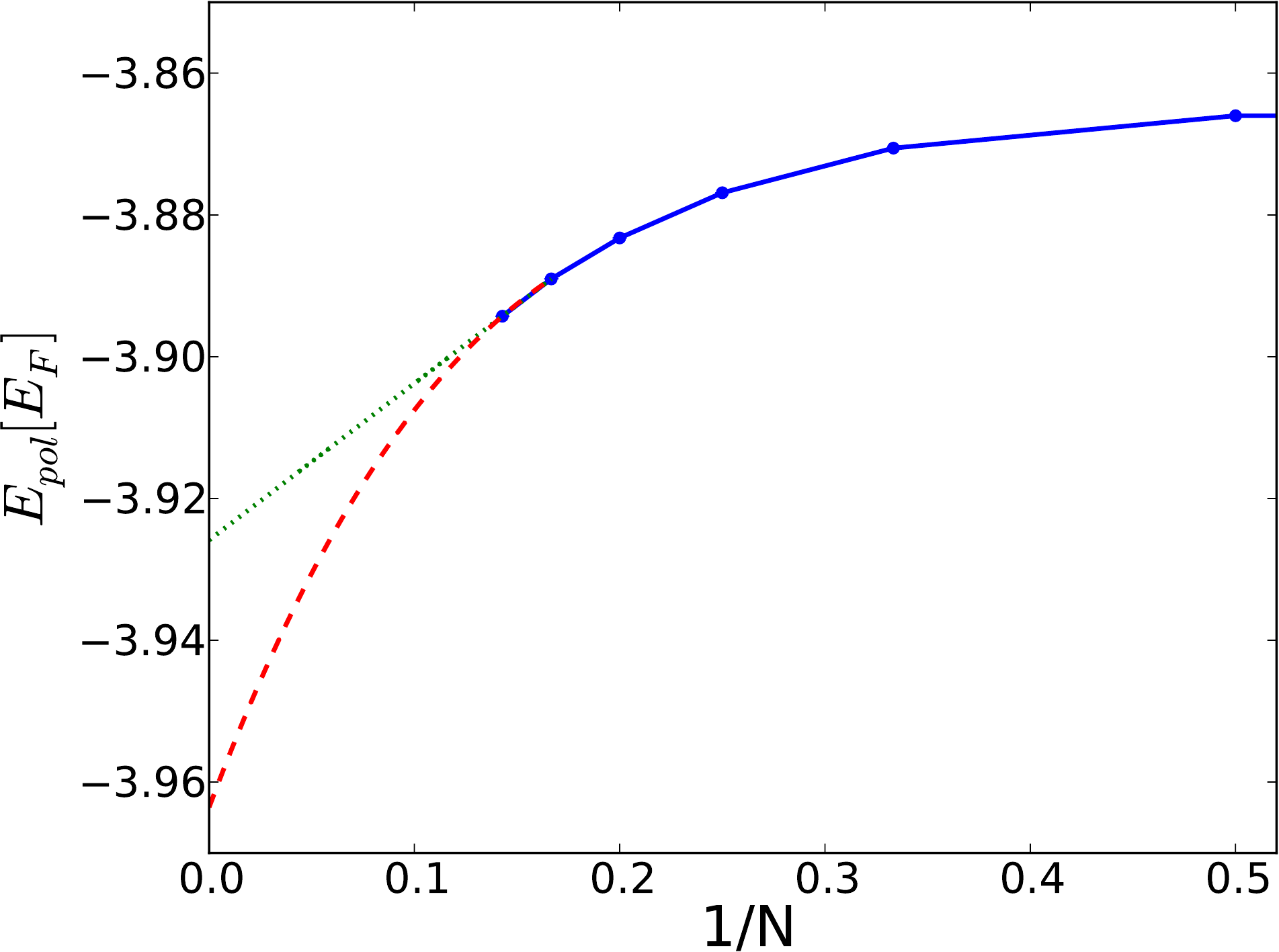}
\caption{\label{fig:resummation1}
(color online) The extrapolation procedure on resummed data is demonstrated for $\eta = -0.248$ and $\delta = 2$.
The dotted curve shows the linear part, whereas the dashed line is a fit with $f$.
}
\end{figure}
\begin{figure}[tbp]
\includegraphics[width=0.99\linewidth]{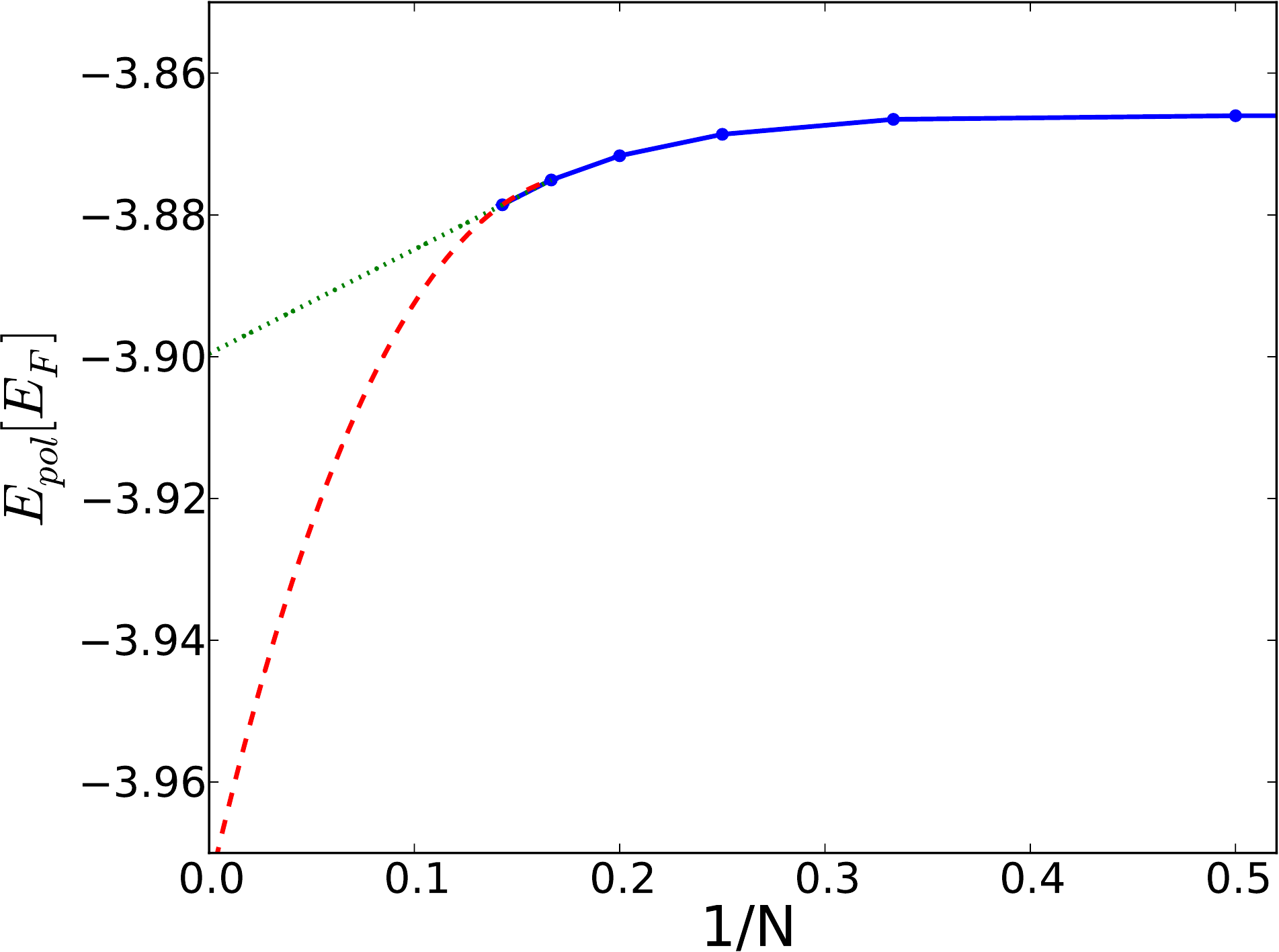}
\caption{\label{fig:resummation2}
(color online) The extrapolation procedure on resummed data is demonstrated for $\eta = -0.248$ and $\delta = 4$.
The dotted curve shows the linear part, whereas the dashed line is a fit with $f$. Note that the resulting error bar 
exceeds the corresponding $\delta = 2$ error bar.
}
\end{figure}
\begin{figure}[tbp]
\includegraphics[width=0.99\linewidth]{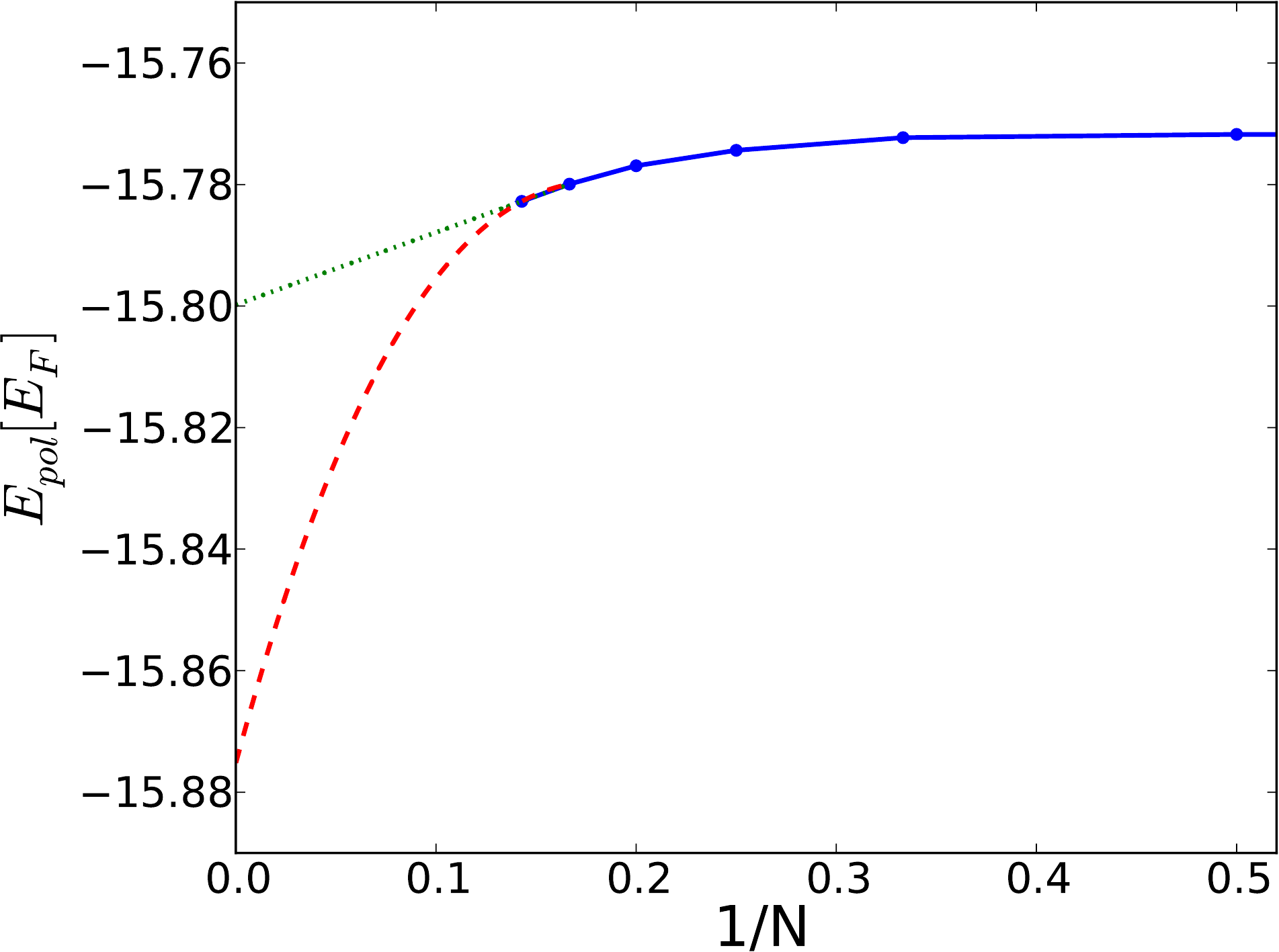}
\caption{\label{fig:resummation3}
(color online) The extrapolation procedure on resummed data is demonstrated for $\eta = -1.02$ and $\delta = 4$.
The dotted curve shows the linear part, whereas the dashed line is a fit with $f$.
}
\end{figure}
\begin{figure}[tbp]
\includegraphics[width=0.99\linewidth]{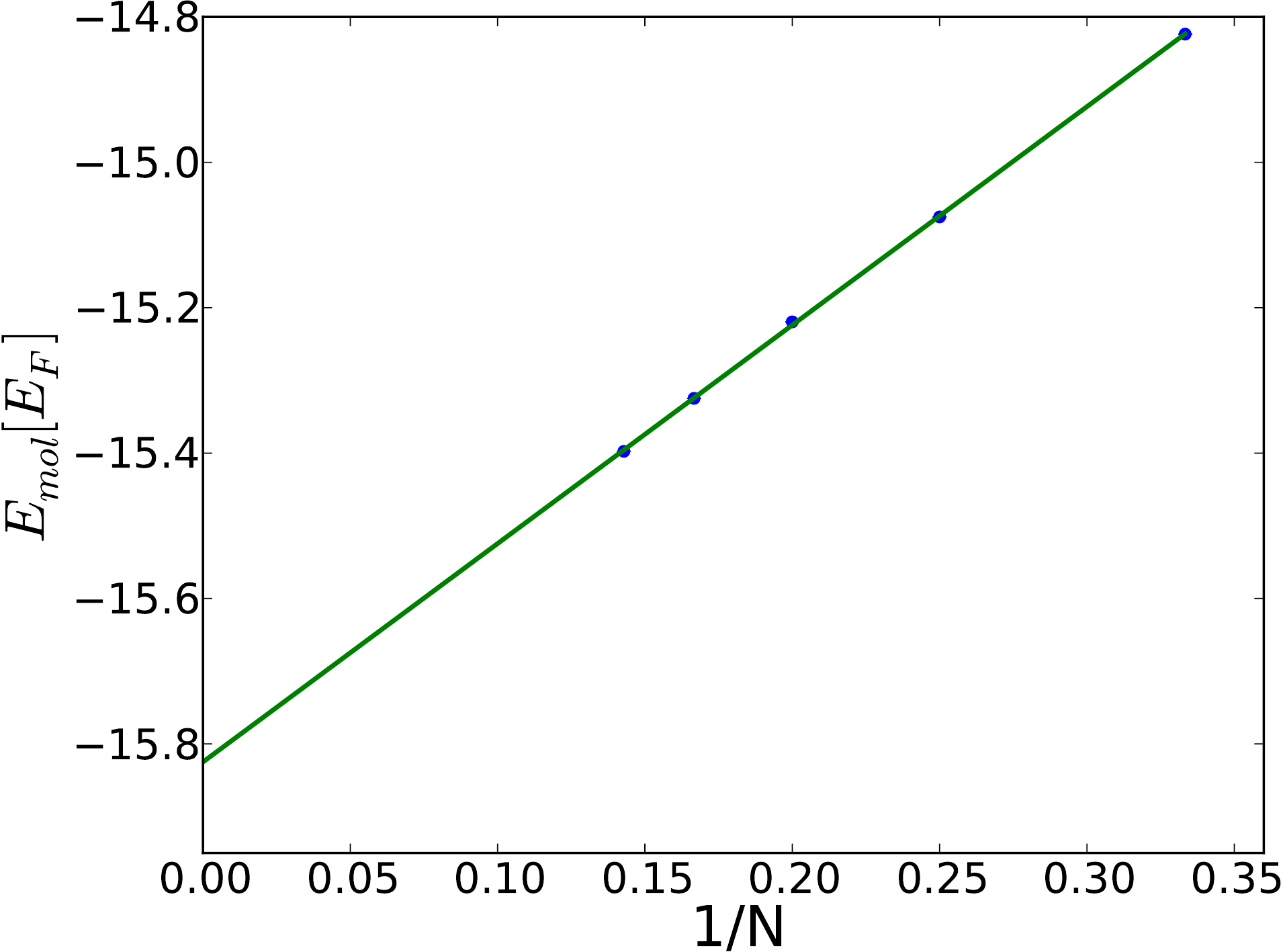}
\caption{\label{fig:resummation4}
(color online) The extrapolation procedure on resummed data is demonstrated for a molecule with $\eta = -1.02$ and $\delta = 6$.
}
\end{figure}
In this appendix, we explain the details of our resummation procedure.
Its use is most delicate for cases where the maximum diagram order is small. Therefore, it is illustrative to use a quasi-two-dimensional
Fermi-polaron series (characterized by the dimensionless parameter $\eta = \ln{(k_F^{\text{2D}} a^{\text{2D}})}$, where $a^{\text{2D}}$ is the
two-dimensional scattering length and $k_F^{\text{2D}}$ is the two-dimensional Fermi momentum) to explain this technique,
since the maximum expansion order is approximately 8.
For these systems, it is additionally necessary to deal with large binding energies, hence aggressive resummation has to be applied to the
bare series in order to be able to extrapolate to infinite expansion order.
However, this tends to conceal the curvature of the series in the first points, leading to an initially flat 
curve.

Our extrapolation procedure is the following:
For the upper value of the error bar, we apply linear extrapolation on the Riesz-resummed data with Riesz exponent
$\delta$. In this linear extrapolation, only the two points corresponding to highest and second highest expansion 
order are taken into account.
For the lower value of the error bar, we {\it assume} a worst-case scenario with large curvature of the extrapolated curve, according
to the following fit function $f$ of parameters $a$ and $b$:
\begin{equation}
\label{fitting}
f(N^{-1},a,b) = 4 \delta \left( \frac{1}{3} N^{-3} - \frac{3}{10} N^{-2} \right) + a N^{-1} + b .
\end{equation}
$N$ denotes the maximum expansion order.
We emphasize that the curvature of this function is empirically set by us. 
This curve includes only the highest and second highest expansion points.
An important feature of $f$ is the dependence on the Riesz exponent. This ensures that a stronger resummation results in a bigger
error bar due to extrapolation errors. The fit $f$ can also be used for bare data ($\delta = 0$). In this case, we replace $\delta$ by -1 
in Eq.~\ref{fitting}. Note that the error bars represent a variability of results due to systematic origins corresponding to the one-$\sigma$-interval.
The result of this technique is shown in Fig.~\ref{fig:resummation1}.
For a maximum expansion order of 7, the two ways of extrapolating are shown in comparison.
If the maximum expansion order was 6, then this error bar would increase, just as one would expect regarding the loss of information.
Fig.~\ref{fig:resummation2} uses a sharper resummation on the same data, demonstrating that the error bar increases with $\delta$.

Therefore, it becomes clear that the weakest possible resummation procedure (among the ones resulting in a monotonously decaying series) should be applied.
As a final example, we show our resummation for a polaron point inside the quasi-two-dimensional transition region in~Fig.~\ref{fig:resummation3}.
Here, resummation with $\delta = 2$ is too weak, so $\delta = 4$ has to be used, resulting in a stronger curvature.
It is important to stress that these two extrapolations represent assumed worst-case scenarios.
Finally, as always for diagMC, the extrapolation result has to be checked with available experimental or
theoretical results, thus justifying its application in retrospect. In our case, these results would be variational 2-ph results
which we expect to be quantitatively exact. As an example for a system in which extrapolated error bars were underestimated
for a similar system, consult Fig.~22 of Ref.~\onlinecite{levinsen2014}.

For molecular energies, resummation is more straightforward. As this series is typically alternating, resummed curves can often
be extrapolated linearly. 
This is demonstrated in Fig.~\ref{fig:resummation4}: The last four points are well fit by a straight line.
However, as this resummation involves the same dangers as described above, we try to vary both the fitting (e.g., fitting three of the
four last points) and the resummation technique to test the variability of this result.


\begin{thebibliography}{99}
\bibitem{order} We define the order of a polaron (molecule) diagram by the number of $T$ matrices (impurity propagators).
\end{thebibliography}
\end{document}